\documentclass[11pt,a4paper]{article}

\usepackage{jcappub}
\usepackage{graphicx}
\usepackage{bm}
\usepackage{latexsym}
\usepackage{psfrag}
\usepackage{ulem}
\usepackage{textcomp}
\usepackage{color}
\usepackage{pstricks}

\makeatletter
\gdef\@fpheader{}
\makeatother

\def\benu{\begin{enumerate}}
\def\eenu{\end{enumerate}}
\def\nn{\nonumber} 

\def\f{\frac}
\def\l{\left}
\def\r{\right}
\def\d{{\rm d}}

\def\cR{{\mathcal R}}

\def\cB{{\cal B}}
\def\cG{{\cal G}}
\def\cI{{\cal I}}
\def\cJ{{\cal J}}
\def\cK{{\cal K}}

\def\ei{\eta_{\rm i}}

\def\ee{\eta_{\rm e}}

\def\vk{{\bm k}}
\def\vka{{\bm k}_{1}}
\def\vkb{{\bm k}_{2}}
\def\vkc{{\bm k}_{3}}
\def\ska{{k_{1}}}
\def\skb{{k_{2}}}
\def\skc{{k_{3}}}
\def\ps{{\mathcal P}_{_{\rm S}}}
\def\pt{{\mathcal P}_{_{\rm T}}}
\def\ns{n_{_{\rm S}}}
\def\nt{n_{_{\rm T}}}
\def\fnl{f_{_{\rm NL}}}
\def\hnl{h_{_{\rm NL}}}
\def\cnls{C_{_{\rm NL}}^{\mathcal R}}
\def\cnlt{C_{_{\rm NL}}^{\mathcal \gamma}}
\def\Mpl{M_{_{\rm Pl}}}
\def\Mp{M_{_{\rm Pl}}}

\def\cRB{{\cal R}^{_{^{^{\rm B}}}}}
\def\cRBe{{\cal R}^{_{{^{\rm B}}}}}
\def\gB{\gamma^{_{\rm B}}}
\def\gBe{\gamma^{_{^{\rm B}}}}

\newcommand{\sr}{{\cal R}}
\newcommand{\g}{\gamma}

\newcommand{\bea}{\begin{eqnarray}}
\newcommand{\eea}{\end{eqnarray}}

\begin{document}
\title{Examining the consistency relations describing the three-point 
functions involving tensors}
\author{V.~Sreenath}
\affiliation{Department of Physics, Indian Institute of 
Technology Madras, Chennai~600036, India}
\emailAdd{sreenath@physics.iitm.ac.in}
\author{and L.~Sriramkumar}
\emailAdd{sriram@physics.iitm.ac.in}
\date{today} 
\abstract{It is well known that the non-Gaussianity parameter $\fnl$ 
characterizing the scalar bi-spectrum can be expressed in terms of the 
scalar spectral index in the squeezed limit, a property that is referred 
to as the consistency relation. 
In contrast to the scalar bi-spectrum, the three-point cross-correlations 
involving scalars and tensors and the tensor bi-spectrum have not received
adequate attention, which can be largely attributed to the fact that the 
tensors had remained undetected at the level of the power spectrum until 
very recently. 
The detection of the imprints of the primordial tensor perturbations by 
BICEP2 and its indication of a rather high tensor-to-scalar ratio, if 
confirmed, can open up a new window for understanding the tensor 
perturbations, not only at the level of the power spectrum, but also in 
the realm of non-Gaussianities.
In this work, we consider the consistency relations associated with the 
three-point cross-correlations involving scalars and tensors as well as 
the tensor bi-spectrum in inflationary models driven by a single, 
canonical, scalar field. 
Characterizing the cross-correlations in terms of the dimensionless 
non-Gaussianity parameters $\cnls$ and $\cnlt$ that we had introduced 
earlier, we express the consistency relations governing the 
cross-correlations as relations between these non-Gaussianity parameters 
and the scalar or tensor spectral indices, in a fashion similar to that 
of the purely scalar case. 
We also discuss the corresponding relation for the non-Gaussianity 
parameter $\hnl$ used to describe the tensor bi-spectrum.
We analytically establish these consistency relations explicitly in the 
following two situations: a simple example involving a specific case of 
power law inflation and a non-trivial scenario in the so-called Starobinsky 
model that is governed by a linear potential with a sharp change in its 
slope.
We also numerically verify the consistency relations in three types of 
inflationary models that permit deviations from slow roll and lead to 
scalar power spectra with features which typically result in an improved 
fit to the data than the more conventional, nearly scale invariant, 
spectra.
We close with a summary of the results we have obtained.}
\maketitle


\section{Introduction}\label{sec:introduction}

Our current understanding of the universe on large scales is based on the 
$\Lambda$CDM model, supplemented by the inflationary paradigm. 
While there exist certain limited alternatives to inflation, none of the 
alternatives seem to perform as effectively against the various cosmological 
data as inflation seems capable of.
The inflationary paradigm was originally proposed to ensure that the initial
conditions for the background cosmological model are more natural than what
was possible within the hot big bang theory.
But, it was soon recognized that the paradigm also provides a simple and 
efficient mechanism for the origin of the perturbations.
Inflation can be achieved rather easily using scalar fields (usually referred 
to as the inflaton), and it is the quantum fluctuations associated with the scalar
fields that sow the seeds of the primordial perturbations~\cite{texts,reviews}.
These perturbations lead to the anisotropies in the Cosmic Microwave Background 
(CMB) and, eventually, to the formation of the large scale structure. 
Despite the measurements of the CMB anisotropies with ever increasing precision 
by missions such as WMAP and Planck, details concerning the dynamics of the 
scalar field driving inflation still remain to be satisfactorily 
understood~\cite{wmap-2011,wmap-2013,planck-2013-cmbps,planck-2013-ccp,planck-2013-ci}.
Specifically, whereas the data seems to point to inflation driven by 
a slowly rolling scalar field, we are rather far from converging on the 
form of the potential governing the inflaton (in this context, see, for 
instance, Refs.~\cite{be-fim,martin-2013-14}).
 
\par

Very often, inflationary models are compared with the cosmological data at 
the level of the scalar power spectrum.
Over the last decade, it has been realized that non-Gaussianities and, in
particular, the scalar bi-spectrum can provide a powerful handle to arrive 
at a much smaller class of viable inflationary models (for the earliest 
efforts in this direction, see Refs.~\cite{earlyng}; for various theoretical 
efforts, see, for example, Refs.~\cite{maldacena-2003,ng-ncsf,ng-reviews};
for work prior to Planck on arriving at constraints on non-Gaussianities from 
observations, see, for instance, Refs.~\cite{ng-da,ng-da-reviews}).
Such an expectation has been corroborated to a substantial extent by the 
strong constraints that have been arrived at from the Planck data on the 
three non-Gaussianity parameters, viz. ($\fnl^{\rm loc}$, $\fnl^{\rm eq}$, 
$\fnl^{\rm ortho}$), that are commonly used to characterize the scalar 
bi-spectrum~\cite{planck-2013-cpng}.
While a considerable amount of effort has been dedicated to understanding 
the generation and imprints of the scalar bi-spectrum, a rather limited 
amount of attention has been paid to investigating the three-point functions 
involving the tensor 
perturbations~\cite{tensor-bs,tanaka-2011,cc,cc2,sreenath-2013,kundu-2013}.
This can be obviously ascribed to the fact that the tensors had remained 
undetected at the level of the power spectrum until very recently.
Needless to add, the detection of the imprints of the primordial tensor 
perturbations on the CMB by BICEP2~\cite{bicep2}, if it is also confirmed 
by, say, the forthcoming polarization data from Planck, would significantly 
alter the situation. 
Importantly, the high tensor-to-scalar ratio implied by the BICEP2 
measurements provides hope that it may not be too far fetched to imagine 
that we can arrive at observational constraints on the three-point 
cross-correlations consisting of scalars and tensors and the tensor 
bi-spectrum in the foreseeable future\footnote{We should hasten to add a 
clarifying remark here. 
A priori, the three-point functions are an independent measure of the
primordial perturbations. 
So, a high tensor-to-scalar ratio does not necessarily imply significant
amplitudes for the three-point functions involving tensors.
But, evidently, detecting, say, the tensor bi-spectrum, seems less likely 
if the tensor-to-scalar ratio actually turns out to be considerably smaller 
than what BICEP2 has observed.}.

\par

In a recent work, we had constructed a procedure (and developed a code)
for numerically evaluating the three-point cross-correlations comprising 
of scalars and tensors as well as the tensor bi-spectrum for an arbitrary 
triangular configuration of the three wavenumbers involved~\cite{sreenath-2013}.
We had also introduced dimensionless non-Gaussianity parameters, which we
had denoted as $\cnls$ and $\cnlt$, to characterize the amplitude of the 
three-point scalar-tensor cross-correlations.   
It is well known that, in the squeezed limit wherein one of the wavenumbers 
is much smaller than the other two, the inflationary scalar bi-spectrum 
generated by a single scalar field can be expressed completely in terms of 
the scalar power spectrum, a result that is referred to as a consistency 
relation (for the original results, see, for instance, 
Refs.~\cite{maldacena-2003,creminelli-2004}; for more recent discussions in 
this context, see Refs.~\cite{cr-rd}; for similar results involving higher 
order correlation functions, see, for example, Refs.~\cite{npfs}).
Or, equivalently, the scalar non-Gaussianity parameter $\fnl$ can be
written purely in terms of the scalar spectral index.
However, most of the work on the consistency relations have focussed on
the scalar bi-spectrum, and it seems natural to expect that similar 
consistency relations will be satisfied by the three-point functions
that involve the tensor perturbations as well (in this context, see 
Refs.~\cite{maldacena-2003,tanaka-2011,cc2,kundu-2013}).
Our aim in this work is to investigate the validity of consistency 
relations involving the three-point scalar-tensor cross-correlations 
and the tensor bi-spectrum.
We shall first express the corresponding consistency relations as relations 
between the non-Gaussianity parameters ($\cnls$ and 
$\cnlt$~\cite{sreenath-2013}, and the quantity $\hnl$ that is used to 
describe the purely tensor case~\cite{tensor-bs}) and the scalar or the 
tensor spectral indices.
We shall then analytically as well as numerically examine the validity
of these consistency conditions in specific situations.  
As we shall illustrate, the consistency relations hold generically,
and they prove to be valid even in scenarios involving substantial 
deviations from slow roll.

\par

The structure of this paper is as follows.
In the following section, we shall quickly introduce the quantities 
characterizing the scalar and tensor perturbations and the standard 
definitions of the power spectra in terms of these quantities.
We shall also discuss the different three-point functions of interest
and introduce the corresponding non-Gaussianity parameters.
We shall further summarize the essential expressions regarding the 
evaluation of the three-point functions in the Maldacena formalism 
in single field inflationary models involving the canonical scalar 
field.
In the subsequent section, we shall outline a proof of the consistency relations
which describe the behavior of the three-point functions in the squeezed limit, 
wherein they can be expressed in terms of the scalar and the tensor power spectra.
We shall also state the consistency relations as relations between the four 
non-Gaussianity parameters, viz. $\fnl$, $\cnls$, $\cnlt$ and $\hnl$, and 
the scalar and the tensor spectral indices. 
In Sec.~\ref{sec:ae}, we shall explicitly establish the consistency relations 
for the three non-Gaussianity parameters $\cnls$, $\cnlt$ and $\hnl$ in two 
analytically tractable examples---firstly, in a simple situation involving a 
specific case of power law inflation and, secondly, in a non-trivial scenario 
in the so-called Starobinsky model that is described by a linear inflaton 
potential with a sudden change in its slope. 
In Sec.~\ref{sec:ne}, we shall numerically investigate the validity of the 
consistency relations involving $\cnls$, $\cnlt$ and $\hnl$ in models which 
permit deviations from slow roll.
We shall consider three types of models which lead to features in scalar 
power spectrum and are known to result in an improved fit to the CMB data, 
and show numerically that the consistency relations hold in each of these
cases.
Finally, in Sec.~\ref{sec:d}, we conclude with a brief discussion of the results. 

\par

A few words on our notations and conventions are in order at this stage 
of our discussion.
We shall work with units such that $\hbar=c=1$ and assume the Planck mass 
to be $\Mp=\l(8\,\pi\, G\r)^{-1/2}$.
We shall adopt the metric signature of $(-,+,+,+)$, and we shall make
use of Latin indices to denote the spatial coordinates (barring $k$, which 
will represent the wavenumber).
The quantities $a$ and $H$ shall represent the scale factor and the Hubble
parameter of the spatially flat, Friedmann-Lema\^{\i}tre-Robertson-Walker 
(or, simply, Friedmann, hereafter) universe.
We shall work in terms of either the cosmic time $t$ or the the conformal 
time $\eta$, and denote differentiation with respect to these quantities 
as an overdot and an overprime, respectively.


\section{Three-point functions and the associated non-Gaussianity 
parameters}\label{sec:tpfs-ngp}

In this section, we shall quickly summarize a few expressions and results
involving the scalar and tensor perturbations and the two and three-point
correlation functions that will be essential for our discussion. 


\subsection{Primordial perturbations and power spectra}

Upon taking into account the scalar perturbation described by the curvature
perturbation $\cR$ and the tensor perturbation characterized by $\g_{ij}$, 
the spatially flat Friedmann metric can be expressed as~\cite{maldacena-2003}
\begin{equation}
\d s^2 = -\d t^2 
+ h_{ij}(t,{\bm x})\; \d {\bm x}^i\, \d {\bm x}^j,\label{eq:metric}
\end{equation}
where the quantity $h_{ij}$ is given by 
\begin{equation}
h_{ij}(t,{\bm x})=a^{2}(t)\; {\rm e}^{2\,{\cal R}(t,{\bm x})}\,
\l[{\rm e}^{\gamma(t,{\bm x})}\r]_{ij}.\label{eq:metric-spatial}
\end{equation}
Recall that, in the inflationary paradigm, the primordial perturbations 
are generated due to quantum fluctuations.
On quantization, the curvature perturbation $\hat{\cR}$ and the tensor 
perturbation $\hat\g_{ij}$ can be written in terms of the scalar and 
tensor Fourier modes, say, $f_k$ and $g_k$, as follows:
\begin{subequations}
\label{eqs:st-m-dc}
\bea 
\hat{\cR}(\eta, {\bf x}) 
&=& \int \frac{\d^{3}{\bm k}}{\l(2\,\pi\r)^{3/2}}\,
\hat{\cR}_{\bm k}(\eta)\, {\rm e}^{i\,{\bm k}\cdot{\bm x}}\nn\\
&=& \int \frac{\d^{3}{\bm k}}{\l(2\,\pi\r)^{3/2}}\,
\l(\hat{a}_{\bm k}\,f_{k}(\eta)\, 
{\rm e}^{i\,{\bm k}\cdot{\bm x}}
+\hat{a}^{\dagger}_{\bf k}\,f^{*}_{k}(\eta)\,
{\rm e}^{-i\,{\bm k}\cdot{\bm x}}\r),\label{eq:s-m-dc}\\
\hat{\gamma}_{ij}(\eta, {\bf x}) 
&=& \int \frac{\d^{3}{\bm k}}{\l(2\,\pi\r)^{3/2}}\,
\hat{\gamma}_{ij}^{\bm k}(\eta)\, {\rm e}^{i\,{\bm k}\cdot{\bm x}}\nn\\
&=& \sum_{s}\int \frac{\d^{3}{\bm k}}{(2\,\pi)^{3/2}}\,
\l(\hat{b}^{s}_{\bm k}\, \varepsilon^{s}_{ij}({\bm k})\,
g_{k}(\eta)\, {\rm e}^{i\,{\bm k}\cdot{\bm x}}
+\hat{b}^{s\dagger}_{\bf k}\,\varepsilon^{s*}_{ij}({\bm k})\, g^{*}_{k}(\eta)\,
{\rm e}^{-i\,{\bm k}\cdot{\bm x}}\r).\label{eq:t-m-dc}
\eea
\end{subequations}
In these decompositions, the pairs of operators $(\hat{a}_{\bm k},
\hat{a}^{\dagger}_{\bm k})$ and $(\hat{b}_{\bm k}^{s},
\hat{b}^{s\dagger}_{\bm k})$ represent the annihilation and 
creation operators corresponding to the scalar and the tensor 
modes associated with the wavevector ${\bm k}$, and they 
satisfy the standard commutation relations.
The quantity $\varepsilon^{s}_{ij}({\bm k})$ represents the polarization 
tensor of the gravitational waves with their helicity being denoted by the 
index~$s$.
The transverse and traceless nature of the gravitational waves leads to
the conditions $\varepsilon^{s}_{ii}({\bm k})=k_{i}\,
\varepsilon_{ij}^s({\bm k})=0$. 
In this paper, we shall work with a normalization such that 
$\varepsilon_{ij}^{r}({\bm k})\,\varepsilon_{ij}^{s*}({\bm k})
=2\;\delta^{rs}$~\cite{maldacena-2003}.

\par

In terms of the Mukhanov-Sasaki variables, viz. $v_k = z\, f_k$ and $u_k = \Mpl\, 
a\, g_k/\sqrt{2}$, where $z=\sqrt{2\, \epsilon_1}\, \Mp\, a$, with $\epsilon_1
=-{\dot H}/H^2$ being the first slow roll parameter, the equations of motion
governing the scalar and the tensor perturbations reduce to
\begin{subequations}
\label{eqs:ms}
\begin{eqnarray}
v_k''+\l(k^2-\f{z''}{z}\r)\, v_k &=& 0,\label{eq:ms-f}\\
u_k''+\l(k^2-\f{a''}{a}\r)\, u_k &=& 0,\label{eq:ms-g}  
\end{eqnarray}
\end{subequations}
respectively. 
The scalar and the tensor power spectra, viz. ${\mathcal P}_{_{\rm S}}(k)$ 
and ${\mathcal P}_{_{\rm T}}(k)$, are defined as follows:
\begin{subequations}
\begin{eqnarray}
\langle\, {\hat \cR}_{\bm k}(\eta)\,
{\hat \cR}_{\bm k'}(\eta)\,\rangle
&=&\f{(2\,\pi)^2}{2\, k^3}\, {\mathcal P}_{_{\rm S}}(k)\;
\delta^{(3)}({\bm k}+{\bm k'}),\label{eq:sps-d}\\
\langle\, {\hat \gamma}_{ij}^{\bm k}(\eta)\,
{\hat \gamma}_{mn}^{\bm k'}(\eta)\,\rangle
&=&\f{(2\,\pi)^2}{2\, k^3}\, \f{\Pi_{ij,mn}^{{\bm k}}}{4}\;
{\mathcal P}_{_{\rm T}}(k)\;
\delta^{(3)}({\bm k}+{\bm k'}),\label{eq:tps-d}
\end{eqnarray}
\end{subequations}
where the expectation values on the left hand sides are to 
be evaluated in the specified initial quantum state of the 
perturbations, and the quantity $\Pi_{ij,mn}^{\vk}$ is given 
by~\cite{cc,tensor-bs,sreenath-2013}
\begin{equation}
\Pi_{ij,mn}^{\vk}
=\sum_{s}\;\varepsilon_{ij}^{s}(\vk)\;
\varepsilon_{mn}^{s\ast}(\vk).
\end{equation}
The vacuum state $\vert 0\rangle$ associated with the quantized 
perturbations is defined as the state that satisfies the conditions 
${\hat a}_{\bm k}\vert 0\rangle=0$ and ${\hat b}_{\bm k}^{s}\vert 
0\rangle=0$ for all ${\bm k}$ and~$s$.
If one assumes the initial state of the perturbations to be 
the vacuum state $\vert 0\rangle$, then, on making use of the 
decompositions~(\ref{eqs:st-m-dc}) in the above definitions, 
the inflationary scalar and tensor power spectra 
${\mathcal P}_{_{\rm S}}(k)$ and ${\mathcal P}_{_{\rm T}}(k)$
can be expressed as 
\begin{subequations}
\label{eqs:pspt}
\begin{eqnarray}
{\mathcal P}_{_{\rm S}}(k)
&=&\f{k^3}{2\, \pi^2}\, \vert f_k\vert^2,\label{eq:sps}\\
{\mathcal P}_{_{\rm T}}(k)
&=&4\;\f{k^3}{2\, \pi^2}\, \vert g_k\vert^2.\label{eq:tps}
\end{eqnarray}
\end{subequations}
The amplitudes $\vert f_k\vert$ and $\vert g_k\vert$ on the right hand
sides of the above expressions are to be evaluated when the modes are 
sufficiently outside the Hubble radius.
It is useful to note here that the scalar and tensor spectral indices 
$\ns$ and $\nt$ are defined as 
\begin{subequations}
\label{eqs:nsnt}
\begin{eqnarray}
\ns&=&1+\f{\d \ln \ps(k)}{\d \ln k},\label{eq:ns}\\ 
\nt &=& \f{\d \ln \pt(k)}{\d \ln k}.\label{eq:nt} 
\end{eqnarray}
\end{subequations}


\subsection{The non-Gaussianity parameters}

The scalar bi-spectrum, the  two scalar-tensor three-point cross-correlations 
and the tensor bi-spectrum in Fourier space, viz. 
$\cB_{\cR\cR\cR}(\vka,\vkb,\vkc)$,
$\cB_{\cR\cR\gamma}^{m_3n_3}(\vka,\vkb,\vkc)$, 
$\cB_{\cR\gamma\gamma}^{m_2n_2m_3 n_3}(\vka,\vkb,\vkc)$ and 
$\cB_{\gamma\gamma\gamma}^{m_1n_1m_2n_2m_3 n_3}(\vka,\vkb,\vkc)$, evaluated
towards the end of inflation at the conformal time, say, $\ee$, are defined 
as\footnote{In this paper, we shall be primarily interested in the three-point 
functions involving the tensor perturbations.
However, at a couple of locations, we shall also briefly touch upon the 
case of the scalar bi-spectrum and the corresponding non-Gaussianity 
parameter $\fnl$ for the sake of completeness.}
\begin{subequations}
\label{eq:tpfs}
\begin{eqnarray}
\langle\, {\hat \cR}_{\vka}(\eta _{\rm e})\, 
{\hat \cR}_{\vkb}(\eta _{\rm e})\,
{\hat \cR}_{\vkc}(\eta _{\rm e})\, \rangle 
&\equiv&\l(2\,\pi\r)^3\, \cB_{\sr\sr\sr}(\vka,\vkb,\vkc)\;
\delta^{(3)}\l(\vka+\vkb+\vkc\r),\label{eq:sss}\\
\langle\, {\hat \cR}_{\vka}(\eta _{\rm e})\, 
{\hat \cR}_{\vkb}(\eta _{\rm e})\,
{\hat \gamma}_{m_3n_3}^{\vkc}(\eta _{\rm e})\, \rangle 
&\equiv&\l(2\,\pi\r)^3\, \cB_{\sr\sr\g}^{m_3n_3}(\vka,\vkb,\vkc)\;
\delta^{(3)}\l(\vka+\vkb+\vkc\r),\label{eq:sst-cc}\\
\langle\, {\hat \cR}_{\vka}(\eta _{\rm e})\, 
{\hat \gamma}_{m_2n_2}^{\vkb}(\eta _{\rm e})\, 
{\hat \gamma}_{m_3n_3}^{\vkc}(\eta _{\rm e})\,\rangle 
&\equiv&\l(2\,\pi\r)^3\, \cB_{\sr\g\g}^{m_2 n_2 m_3 n_3}(\vka,\vkb,\vkc)\;
\delta^{(3)}\l(\vka+\vkb+\vkc\r),\qquad\quad\label{eq:tts-cc}\\
\langle\, {\hat \gamma}_{m_1n_1}^{\vka}(\eta _{\rm e})\,
{\hat \gamma}_{m_2n_2}^{\vkb}(\eta _{\rm e})\, 
{\hat \gamma}_{m_3n_3}^{\vkc}(\eta _{\rm e})\,\rangle 
&\equiv& \l(2\,\pi\r)^3\, \cB_{\g\g\g}^{m_1n_1m_2 n_2 m_3 n_3}(\vka,\vkb,\vkc)\;
\delta^{(3)}\l(\vka+\vkb+\vkc\r).\label{eq:ttt}\nn\\
\end{eqnarray}
\end{subequations}
For convenience, hereafter, we shall write these correlators as
\begin{equation}
\cB_{\sf ABC}(\vka,\vkb,\vkc)
= \l(2\,\pi\r)^{-9/2}\, G_{\sf ABC}(\vka,\vkb,\vkc),\label{eq:GABC}
\end{equation}
where ${\sf A}$, ${\sf B}$ and ${\sf C}$ refer to either ${\cal R}$ 
or $\gamma$.

\par

As we have discussed before, in the purely scalar case, it proves to 
be convenient to express the level of non-Gaussianity in terms of the  
quantity $\fnl$, which is a suitable dimensionless ratio of the three 
and the two-point functions.
In an analogous manner, one can consider similar dimensionless parameters to
describe the scalar-scalar-tensor and the scalar-tensor-tensor correlators as
well as the tensor bi-spectrum.
Let us denote these non-Gaussianity parameters as $\cnls$, $\cnlt$ and $\hnl$, 
respectively. 
These can be introduced through the following expressions for the curvature and 
the tensor perturbations~\cite{sreenath-2013}:
\begin{subequations}
\begin{eqnarray}
\cR(\eta, {\bm x}) 
&=& \cR^{\rm G}(\eta, {\bm x}) 
- \f{3\,\fnl}{5}\;[{\cal R}^{\rm G}(\eta, {\bm x})]^2 
- \cnls\; \cR^{\rm G}(\eta,{\bm x})\; 
\gamma_{{\bar m}{\bar n}}^{\rm G}(\eta, {\bm x}),\label{eq:cnlr}\\
\gamma_{ij}(\eta, {\bm x}) 
&=& \gamma_{ij}^{\rm G}(\eta, {\bm x}) 
- \hnl\; \gamma_{ij}^{\rm G}(\eta,{\bm x})\; 
\gamma_{{\bar m}{\bar n}}^{\rm G}(\eta,{\bm x})
- \cnlt\; \gamma_{ij}^{\rm G}(\eta,{\bm x})\;
\cR^{\rm G}(\eta,{\bm x}),\label{eq:cnlg}
\end{eqnarray}
\end{subequations}
where $\cR^{\rm G}$ and $\gamma_{ij}^{\rm G}$ denote the Gaussian 
quantities.
In these relations, the overbars that appear on the indices of the Gaussian 
tensor perturbations imply that the indices should be summed over all allowed 
values (for further clarifications, see Ref.~\cite{sreenath-2013}). 
Upon using the above definitions and the Wick's theorem that is
applicable to Gaussian random variables, we find that the four 
non-Gaussianity parameters, viz. $\fnl$, $\cnls$, $\cnlt$ 
and $h_{_{\rm NL}}$, can be expressed in terms of the corresponding
three-point functions and the scalar and the tensor power spectra as 
\begin{subequations}
\label{eqs:ngp}
\begin{eqnarray} 
\fnl(\vka,\vkb,\vkc)
&=&-\frac{5/6}{\l(2\,\pi^2\r)^2}\; 
\l[k_1^3\, k_2^3\, k_3^3\;  G_{\cR\cR\cR}(\vka,\vkb,\vkc)\r]\nn\\
& &\times\l[k_1^{3}\; {\cal P}_{_{\rm S}}(k_2)\; {\cal P}_{_{\rm S}}(k_3)
+{\rm two~permutations}\r]^{-1},\label{eq:fnl}\\
\cnls(\vka,\vkb,\vkc) 
&=& -\f{4}{\l(2\,\pi^2\r)^2}\,
\l[k_1^3\, k_2^3\, k_3^3\; G_{\cR\cR\g}^{m_3n_3}(\vka,\vkb,\vkc)\r]\nn\\
& &\times\; {\l(\Pi_{m_3n_3,{\bar m}{\bar n}}^{\vkc}\r)}^{-1}\;\,
\biggl\{\l[k_1^3\; {\mathcal P}_{_{\rm S}}(k_2)
+k_2^3\; {\mathcal P}_{_{\rm S}}(k_1)\r]\; 
{\mathcal P}_{_{\rm T}}(k_3)\biggr\}^{-1},
\label{eq:cnls}\\
\cnlt(\vka,\vkb,\vkc) 
&=& -\f{4}{\l(2\,\pi^2\r)^2}\,
\l[k_1^3\, k_2^3\, k_3^3\; G_{\cR\g\g}^{m_2n_2m_3n_3}(\vka,\vkb,\vkc)\r]\nn\\
& &\times\,
\biggl\{{\mathcal P}_{_{\rm S}}(k_1)\;
\l[\Pi_{m_2n_2,m_3n_3}^{\vkb}\;k_3^3\; {\mathcal P}_{_{\rm T}}(k_2)
+\Pi_{m_3n_3,m_2n_2}^{\vkc}\; k_2^3\;
{\mathcal P}_{_{\rm T}}(k_3)\r]\biggr\}^{-1},\quad\qquad\label{eq:cnlt}\\
\hnl(\vka,\vkb,\vkc)
&=&-\f{4^2}{\l(2\,\pi^2\r)^2}\,
\l[k_1^3\, k_2^3\, k_3^3\; 
G_{\g\g\g}^{m_1n_1m_2n_2m_3n_3}(\vka,\vkb,\vkc)\r]\nn\\
& &\times\; 
\l[\Pi_{m_1n_1,m_3n_3}^{\vka}\,\Pi_{m_2n_2,{\bar m}{\bar n}}^{\vkb}\;
k_3^3\; {\mathcal P}_{_{\rm T}}(k_1)\;{\mathcal P}_{_{\rm T}}(k_2)
+{\rm five~permutations}\r]^{-1}.\label{eq:hnl} \nn\\
\end{eqnarray}
\end{subequations}


\subsection{The inflationary three-point functions in the Maldacena 
formalism}\label{subsec:tpfs-mf}

The most complete approach to evaluate the three-point functions 
generated during inflation is the formalism originally due to 
Maldacena~\cite{maldacena-2003}. 
The first step in the approach is to obtain the action describing 
the perturbations at the third order.  
Having obtained the third order action describing the perturbations, 
the different three-point correlation functions can be arrived at 
using the standard rules of perturbative quantum field 
theory~\cite{maldacena-2003,cc,tensor-bs,sreenath-2013,kundu-2013}.

\par 

In this work, as we have discussed, we shall be interested in the
three-point functions that involve the tensor perturbations. 
One can show that the scalar-scalar-tensor cross-correlation 
$G_{\sr\sr\gamma}^{m_3n_3}(\vka,\vkb,\vkc)$, when evaluated 
in the perturbative vacuum, can be written as (see, 
for example, Ref.~\cite{sreenath-2013})
\bea \label{eq:Grrg}
G_{\sr\sr\gamma}^{m_3n_3}(\vka,\vkb,\vkc)
& = & \sum_{C=1}^{3}\; G_{\sr\sr\gamma\,(C)}^{m_3n_3}(\vka,\vkb,\vkc)\nn\\
& = & \Mp^2\; \Pi_{m_3n_3,ij}^{\vkc}\, {\hat n}_{1i}\, {\hat n}_{2j}\; 
\sum_{C=1}^{3}\, \bigl[f_{\ska}(\ee)\, f_{\skb}(\ee)\, g_{\skc}(\ee)\nn\\
& & \times\;\cG_{\sr\sr\gamma}^{C}(\vka,\vkb,\vkc)
+\, {\rm complex~conjugate}\bigr],\qquad
\end{eqnarray}
where the quantities $\cG_{\sr\sr\gamma}^{C}(\vka,\vkb,\vkc)$ are described 
by the integrals
\begin{subequations}\label{eqs:cGrrg}
\begin{eqnarray}
\cG_{\sr\sr\gamma}^{1}(\vka,\vkb,\vkc)
&=&-2\, i\; k_1\, k_2\, \int_{\ei}^{\ee} \d\eta\; a^2\, \epsilon_1\, f_{\ska}^{\ast}\,
f_{\skb}^{\ast}\,g_{\skc}^{\ast},\label{eq:cGrrg1}\\
\cG_{\sr\sr\gamma}^{2}(\vka,\vkb,\vkc)
&=&\f{i}{2}\; \f{k_3^2}{k_1\,k_2}\,
\int_{\ei}^{\ee} \d\eta\; a^2\, \epsilon_1^2\, f_{\ska}^{\prime\ast}\,
f_{\skb}^{\prime\ast}\,g_{\skc}^{\ast},\label{eq:cGrrg2}\\
\cG_{\sr\sr\gamma}^{3}(\vka,\vkb,\vkc)
&=&\f{i}{2}\; \f{1}{k_1\,k_2}\,
\int_{\ei}^{\ee} \d\eta\; a^2\, \epsilon_1^2\, 
\l[k_1^2\,f_{\ska}^{\ast}\,f_{\skb}^{\prime\ast}
+k_2^2\,f_{\ska}^{\prime\ast}\,f_{\skb}^{\ast}\r]\,
g_{\skc}^{\prime\ast}.\label{eq:cGrrg3} 
\end{eqnarray}
\end{subequations}
The lower limit of the integrals, viz. $\ei$, denotes a sufficiently 
early time at which the initial conditions are imposed on the modes 
when they are well inside the Hubble radius.
The upper limit $\ee$ denotes a suitably late time which can, for 
instance, be conveniently chosen to be a time close to the end of 
inflation.
Note that, for a given wavevector ${\bm k}$, ${\hat {\bm n}}$ denotes 
the unit vector ${\hat {\bm n}}={\bm k}/k$.
Hence, the quantities ${\hat n}_{1i}$ and ${\hat n}_{2i}$ represent 
the components of the unit vectors ${\hat {\bm n}}_{1}={\bm k}_1/k_1$ 
and ${\hat {\bm n}}_{2}={\bm k}_2/k_2$ along the $i$-spatial direction.

\par

Similarly, the scalar-tensor-tensor cross-correlation 
$G_{\cR\gamma\gamma}^{m_2n_2m_3 n_3}(\vka,\vkb,\vkc)$,
evaluated in the perturbative vacuum, can be expressed as
\begin{eqnarray} 
G_{\sr\gamma\gamma}^{m_2n_2m_3n_3}(\vka,\vkb,\vkc)
&= & \sum_{C=1}^{3}\;
G_{\sr\gamma\gamma\,(C)}^{m_2n_2m_3n_3}(\vka,\vkb,\vkc)\nn\\
&= & \Mp^2\; \Pi_{m_2n_2,ij}^{\vkb}\; \Pi_{m_3n_3,ij}^{\vkc}\;
\sum_{C=1}^{3}\; \bigl[f_{\ska}(\ee)\, g_{\skb}(\ee)\,g_{\skc}(\ee)\nn\\
& &\times\; \cG_{\sr\gamma\gamma}^{C}(\vka,\vkb,\vkc) 
+ {\rm complex~conjugate}\bigr],\label{eq:Grgg}
\end{eqnarray}
with the quantities $\cG_{\sr\g\g}^{C}(\vka,\vkb,\vkc)$ being given by 
\begin{subequations}
\label{eqs:cGrgg}
\begin{eqnarray}
\cG_{\sr\g\g}^{1}(\vka,\vkb,\vkc)
&=&\f{i}{4}\, \int_{\ei}^{\ee} \d\eta\; a^2\, \epsilon_1\, f_{\ska}^{\ast}\,
g_{\skb}^{\prime\ast}\,g_{\skc}^{\prime\ast},\label{eq:cGrgg1}\hspace{1.4in}\\
\cG_{\sr\g\g}^{2}(\vka,\vkb,\vkc)
&=&-\f{i}{4}\,  \l(\vkb\cdot\vkc\r)\,
\int_{\ei}^{\ee} \d\eta\; a^2\, \epsilon_1\, f_{\ska}^{\ast}\,
g_{\skb}^{\ast}\,g_{\skc}^{\ast},\label{eq:cGrgg2}\\
\cG_{\sr\g\g}^{3}(\vka,\vkb,\vkc)
&=&-\f{i}{4}\, \int_{\ei}^{\ee} \d\eta\; a^2\, \epsilon_1\, 
f_{\ska}^{\prime\ast}\,
\biggl[\f{\vka\cdot\vkb}{k_1^2}\,g_{\skb}^{\ast}\,
g_{\skc}^{\prime\ast}
+\f{\vka\cdot\vkc}{k_1^2}\,\,g_{\skb}^{\prime\ast}\,
g_{\skc}^{\ast}\biggr].\label{eq:cGrgg3}
\end{eqnarray}
\end{subequations}
Lastly, the tensor bi-spectrum 
$G_{\gamma\gamma\gamma}^{m_1n_1m_2n_2m_3n_3}(\vka,\vkb,\vkc)$,
calculated in the perturbative vacuum, can be written 
as~\cite{maldacena-2003,cc,tensor-bs,sreenath-2013,kundu-2013}
\bea \label{eq:Gggg}
G_{\gamma\gamma\gamma}^{m_1n_1m_2n_2m_3n_3}(\vka,\vkb,\vkc)
&= & \Mp^2\; \biggl[\bigl(\Pi_{m_1n_1,ij}^{\vka}\,\Pi_{m_2n_2,im}^{\vkb}\,
\Pi_{m_3n_3,lj}^{\vkc}\nn\\
& &-\,\f{1}{2}\;\Pi_{m_1n_1,ij}^{\vka}\,\Pi_{m_2n_2,ml}^{\vkb}\,
\Pi_{m_3n_3,ij}^{\vkc}\bigr)\, k_{1m}\, k_{1l}
+{\rm five~permutations}\biggr]\nn\\
& &\times\; \bigl[g_{\ska}(\ee)\, g_{\skb}(\ee)\,g_{\skc}(\ee)\nn\\
& &\times\,\cG_{\gamma\gamma\gamma}(\vka,\vkb,\vkc)
+ {\rm complex~conjugate}\bigr],
\end{eqnarray}
where the quantity $\cG_{\gamma\gamma\gamma}(\vka,\vkb,\vkc)$ is
described by the integral
\begin{equation}
\cG_{\g\g\g}^{1}(\vka,\vkb,\vkc)
=-\f{i}{4}\,\int_{\ei}^{\ee} \d\eta\; a^2\, g_{\ska}^{\ast}\,
g_{\skb}^{\ast}\,g_{\skc}^{\ast},\label{eq:cGggg}
\end{equation}
and we should emphasize that $(k_{1i},k_{2i}, k_{3i})$ denote the 
components of the three wavevectors $({\bm k}_1, {\bm k}_{2}, {\bm k}_{3})$ 
along the $i$-spatial direction\footnote{Such an emphasis seems
essential to avoid confusion between $k_1$, $k_2$ and $k_3$ which
denote the wavenumbers associated with the wavevectors ${\bm k}_1$, 
${\bm k}_{2}$ and ${\bm k}_{3}$, and the quantity $k_i$ which 
represents the component of the wavevector ${\bm k}$ along the 
$i$-spatial direction. 
We have made similar clarifications below wherever we are concerned 
some confusion in the notation may arise.}.

\par

The delta functions that appear in the definitions~(\ref{eq:tpfs}) of 
the three-point functions imply that the wavevectors $\vka$, $\vkb$ and 
$\vkc$ form a triangle.
The squeezed limit of the three-point functions corresponds to the 
situation wherein one of the three wavenumbers, i.e. $k_1$, $k_2$ 
or $k_3$, vanishes. 
In the two cases of the scalar-tensor cross-correlations, the squeezed 
limit can evidently be arrived at by choosing the wavenumber of either 
the scalar or the tensor modes to be zero. 
However, the contributions to the scalar-scalar-tensor three-point function
either explicitly involve the wavenumber of the scalar mode or its time 
derivative, both of which go to zero in the large scale limit.
As a result, the scalar-scalar-tensor three-point function itself vanishes
in the large scale limit of either of the scalar modes. 
For the same reasons, one finds that the scalar-tensor-tensor cross-correlation 
also vanishes in the limit wherein the wavenumber of any of the two tensor
modes goes to zero.
Therefore, in order to understand the behavior in the squeezed limit, 
we shall consider the large scale limits of the tensor and the scalar
modes in the cases of the scalar-scalar-tensor and the 
scalar-tensor-tensor cross-correlations, respectively.
In the squeezed limit, the expressions for the cross-correlations 
$G_{\sr\sr\gamma}^{m_3n_3}(\vka, \vkb, \vkc)$ and
$G_{\sr\gamma\gamma}^{m_2n_2m_3n_3}(\vka,\vkb,\vkc)$, and the tensor 
bi-spectrum $G_{\gamma\gamma\gamma}^{m_1n_1m_2n_2m_3n_3}(\vka,\vkb,\vkc)$
can be written as
\begin{subequations} 
\begin{eqnarray} 
\lim_{k_3 \rightarrow 0}\, G_{\sr\sr\gamma}^{m_3n_3}(\vka, \vkb, \vkc)
& = & \lim_{k_3 \rightarrow 0} 
\sum_{C=1}^{3}\; G_{\sr\sr\gamma\,(C)}^{m_3n_3}(\vk,-\vk,\vkc)\nn\\
& = & -\lim_{k_3 \rightarrow 0}\, \Mp^2\; 
\Pi_{m_3n_3,ij}^{\vkc}\, {\hat n}_{i}\, {\hat n}_{j}\; 
\sum_{C=1}^{3}\, \bigl[f_{k}(\ee)\, f_{k}(\ee)\, g_{k_3}(\ee)\nn\\
& & \times\;\cG_{\sr\sr\gamma}^{C}(\vk,-\vk,\vkc)
+\, {\rm complex~conjugate}\bigr],\label{eq:Grrg-sl}\\ 
\lim_{k_1 \rightarrow 0}\, G_{\sr\gamma\gamma}^{m_2n_2m_3n_3}(\vka,\vkb,\vkc)
&= & \lim_{k_1 \rightarrow 0} \sum_{C=1}^{3}\;
G_{\sr\gamma\gamma\,(C)}^{m_2n_2m_3n_3}(\vka,\vk,-\vk)\nn\\
&= & \lim_{k_1 \rightarrow 0}\,
\Mp^2\; \Pi_{m_2n_2,ij}^{\vk}\; \Pi_{m_3n_3,ij}^{-\vk}\;
\sum_{C=1}^{3}\; \bigl[f_{\ska}(\ee)\, g_{k}(\ee)\,g_{k}(\ee)\nn\\
& &\times\; \cG_{\sr\gamma\gamma}^{C}(\vka,\vk,-\vk) 
+ {\rm complex~conjugate}\bigr]\nonumber\\
&=& \lim_{k_1 \rightarrow 0}\, 2\,\Mp^2\; \Pi_{m_2n_2,m_3n_3}^{\vk}\;
\sum_{C=1}^{3}\; \bigl[f_{\ska}(\ee)\, g_{k}(\ee)\,g_{k}(\ee)\nn\\
& &\times\; \cG_{\sr\gamma\gamma}^{C}(\vka,\vk,-\vk) 
+ {\rm complex~conjugate}\bigr],\label{eq:Grgg-sl}\\
\lim_{k_3 \rightarrow 0}\, 
G_{\gamma\gamma\gamma}^{m_1n_1m_2n_2m_3n_3}(\vka,\vkb,\vkc) 
&=&\lim_{k_3 \rightarrow 0}\, 
G_{\gamma\gamma\gamma}^{m_1n_1m_2n_2m_3n_3}(\vk,-\vk,\vkc)\nn\\
&=&-\lim_{k_3 \rightarrow 0}\, \Mp^2\; \Pi_{m_1n_1,ij}^{\vk}\,
\Pi_{m_2n_2,ij}^{-\vk}\, \Pi_{m_3n_3,ml}^{\vkc}\, k_{m}\, k_{l}\nn\\
& &\times\; \bigl[g_{k}(\ee)\, g_{k}(\ee)\,g_{\skc}(\ee)\,
\cG_{\gamma\gamma\gamma}(\vk,-\vk,\vkc)\nn\\
& &+\, {\rm complex~conjugate}\bigr] \nn\\
&=&-\lim_{k_3 \rightarrow 0}\,  2\,\Mp^2\; \Pi_{m_1n_1,m_2n_2}^{\vk}\,
\Pi_{m_3n_3,ml}^{\vkc}\, k_{m}\, k_{l}\nn\\
& &\times\; \bigl[g_{k}(\ee)\, g_{k}(\ee)\,g_{\skc}(\ee)\, 
\cG_{\gamma\gamma\gamma}(\vk,-\vk,\vkc) \nn\\
& & +\, {\rm complex~conjugate}\bigr],\label{eq:Gggg-sl}
\end{eqnarray}
\end{subequations}
where, for simplicity, we have set $\vka=-\vkb=\vk$ in the first 
and the third expressions, and $\vkb=-\vkc=\vk$ in the second.
The overall minus sign in the above scalar-scalar-tensor correlation
arises due to the fact that, in the squeezed limit, ${\hat n}_{1i}=
-{\hat n}_{2i}={\hat n}_{i}$.  
The polarization factors in the tensor bi-spectrum simplify in the squeezed 
limit due to the transverse nature of the gravitational waves, i.e. $k_{i}\; 
\varepsilon_{ij}^s({\bm k})=0$.
Moreover, it is the normalization of the polarization tensor, viz. 
$\varepsilon_{ij}^{r}({\bm k})\;\varepsilon_{ij}^{s*}({\bm k})
=2\;\delta^{rs}$, that leads to the overall factor of two in the cases of
the scalar-tensor-tensor correlation and the tensor bi-spectrum.
Note that the two three-point cross-correlations contain three independent 
terms.  
It is straightforward to argue that, in the squeezed limit of the 
tensor mode, it is only the first term that contributes in the case of the
scalar-scalar-tensor correlation (as the other two contributions either 
depend explicitly on the wavenumber of the squeezed mode or its time 
derivative).  
Similarly, in the case of the scalar-tensor-tensor correlation, one finds 
that the third term does not contribute in the squeezed limit of the scalar 
mode.

\par

In the next section, we shall briefly outline a proof of the consistency 
relations obeyed by the different three-point functions and the 
corresponding non-Gaussianity parameters in the squeezed limit.
 

\section{Consistency relations in the squeezed limit}\label{sec:cr-sl}

A consistency relation basically links the three-point function to 
the two-point function in a particular limit of the wavenumbers 
involved\footnote{In fact, depending on the symmetries associated with 
the action governing the field(s) of interest, quantum field theory suggests
such relations can exist between generic $N$-point and $(N\!-\!1)$-point 
correlation functions.
Clearly, similar connections can be expected to arise for the various
correlation functions generated during inflation as well (in this 
context, see, for instance, Refs.~\cite{npfs}).}.
It has been known for a while that the scalar bi-spectrum obeys a consistency 
relation in the squeezed limit~\cite{maldacena-2003,creminelli-2004,cr-rd}. 
In terms of the scalar non-Gaussianity parameter $\fnl$, it can be 
expressed as $\fnl = 5\; (\ns -1)/12$, where $\ns$ is the scalar 
spectral index [cf. Eq.~(\ref{eqs:nsnt})]. 
The scalar consistency relation is expected to be valid for any single 
field inflationary model, irrespective of the detailed dynamics of the
field~\cite{creminelli-2004}.
As we shall discuss in some detail below, this essentially occurs 
because of the fact that the amplitude of the long wavelength scalar 
modes freezes on super-Hubble scales. 
Due to this reason, their effects on the smaller wavelength modes can 
be treated as though they are evolving in a background with modified
spatial coordinates.   
Since the tensor modes too behave in the same fashion as the scalar 
modes when they are sufficiently outside the Hubble radius (i.e. their 
amplitudes freeze as well), it seems natural to expect 
that there should exist similar consistency relations describing
the three-point scalar-tensor cross-correlations and the tensor 
bi-spectrum~\cite{cc2,tanaka-2011,kundu-2013}. 
In the remainder of this section, we shall arrive at the consistency 
relations governing all the four three-point functions in the squeezed
limit.

\par

As we have already pointed out, the amplitude of a long wavelength scalar 
or tensor mode would be a constant, since they would be well outside the 
Hubble radius (in this context, see, for instance, Refs.~\cite{e-dfsr}).
Due to the fact that the amplitude has frozen, they can be treated as a
background as far as the smaller wavelength modes are concerned.
Let us denote the constant amplitude (i.e. as far as their time dependence 
is concerned) of the long wavelength scalar and tensor modes as, say, $\cRB$ 
and $\gB_{ij}$, respectively. 
In the presence of such modes, the {\it background}\/ metric will take the 
form
\begin{equation}
\d s^2 = -\d t^2 + a^2(t)\; {\rm e}^{2\,\cRBe}\;
[{\rm e}^{\gBe}]_{ij}\; \d {\bm x}^i\, \d {\bm x}^j,
\end{equation}
i.e. the long wavelength modes lead to modified spatial coordinates.
Such a modification is, in fact, completely equivalent to a spatial 
transformation of the form ${\bm x'} = \Lambda\, {\bm x}$, with the
components of the matrix $\Lambda$ being given by $\Lambda_{ij}
={\rm e}^{\cRBe}\, [{\rm e}^{\gBe/2}]_{ij}$. 
Under such a spatial coordinate transformation, one can easily show that 
the Fourier modes of the small wavelength scalar and tensor perturbations 
transform as follows:~${\cal R}_\vk \to {\rm det}~(\Lambda^{-1})\; 
\cR_{\Lambda^{-1}\, \vk}$ and $\g^\vk_{ij} \to {\rm det}~(\Lambda^{-1})\; 
\g^{\Lambda^{-1}\, \vk}_{ij}$, where, evidently, $\Lambda^{-1}$ represents
the inverse of the original matrix $\Lambda$.
Upon using the property that the determinant of the exponential of a 
matrix is the exponential of its trace and the fact that $\gamma_{ij}$ 
is traceless, one arrives at the result ${\det}~(\Lambda^{-1})= 
{\rm e}^{-3\, \cRBe}$.  
At the leading order in $\cRB$ and $\gB$, one can also obtain that
$\vert\Lambda^{-1}\, \vk\vert=[1-\cRB- \gB_{ij}\,k_i\, k_j/(2\,k^2)]\; k$,
where, as we have clarified earlier, $k_i$ denotes the component 
of the wavevector ${\bm k}$ along the $i$-spatial direction.  
Moreover, since $\delta^{(3)}(\Lambda^{-1}\,{\bm k}_1+\Lambda^{-1}\,
{\bm k}_2)={\rm det}~(\Lambda)\; \delta^{(3)}({\bm k}_1+{\bm k}_2)$, 
on combining the above results, one finds that the scalar and the 
tensor two-point functions in the presence of a long wavelength mode 
denoted by, say, the wavenumber $k$, can be written as 
\begin{subequations}
\begin{eqnarray}
\langle \hat{\cR}_{\vka}\, \hat{\cR}_{\vkb} \rangle_{k} 
&=& \f{(2\,\pi)^2}{2\,k_1^3}\; \ps(k_1)\; \delta^{(3)}(\vka + \vkb)\nn\\
& &\times\,\left[1 - (\ns-1)\, \cRB - \l(\frac{\ns -4}{2}\r)\,  
\gB_{ij}\,{\hat n}_{1i}\,{\hat n}_{1j}\r],\\
\langle \hat{\g}^{\vka}_{m_1n_1}\, \hat{\g}^{\vkb}_{m_2n_2}\rangle_{k} 
&=& \f{(2\,\pi)^2}{2\,k_1^3}\; 
\frac{\Pi^{\vka}_{m_1n_1,m_2n_2}}{4}\; \pt(k_1)\; 
\delta^{(3)}(\vka + \vkb)\nn\\
& &\times\,\l[ 1 - \nt\, \cRB - \l(\frac{\nt-3}{2}\r)\,
\gB_{ij}\, {\hat n}_{1i}\,{\hat n}_{1j}\r],
\end{eqnarray}
\end{subequations}
where, as we have mentioned, ${\hat n}_{1i}=k_{1i}/k_1$.

\par

The above expressions for the two-point functions can then be utilized 
to arrive at the four three-point functions in the squeezed limit.
We find that, in the presence of a long wavelength mode, the three-point 
functions can be obtained to be
\begin{subequations}
\begin{eqnarray}
\langle\, \hat{\cR}_{\vka}\, \hat{\cR}_{\vkb}\, 
\hat{\cR}_{\vkc}\, \rangle_{k_3} 
&\equiv&  \langle\, \langle\, \hat{\cal R}_{\vka}\, \hat{\cal R}_{\vkb}\, 
\rangle_{k_3}\; \hat{\cal R}_{\vkc}\, \rangle \nn \\
&=& -\,\f{(2\,\pi)^{5/2}}{4\, k_1^3\, k_3^3}\, \l(\ns - 1\r)\,
\ps(k_1)\, \ps(k_3)\; \delta^{3}(\vka + \vkb),\\  
\langle\, \hat{\cal R}_{\vka}\, \hat{\cal R}_{\vkb}\, 
\hat{\g}^{\vkc}_{m_3n_3}\, \rangle_{k_3} 
&\equiv& 
\langle \, \langle \hat{\cal R}_{\vka}\, \hat{\cal R}_{\vkb}\, \rangle_{k_3}\;
\hat{\g}^{\vkc}_{m_3n_3}\, \rangle\nn\\
&=& -\,\f{(2\,\pi)^{5/2}}{4\, k_1^3\, k_3^3}\, 
\l(\f{\ns - 4}{8}\r)\, \ps(k_1)\, \pt(k_3)\nn\\
& &\times\,\Pi^{\vkc}_{m_3n_3,ij}\, {\hat n}_{1i}\, {\hat n}_{1j}\;
\delta^{3}(\vka+ \vkb),\\ 
\langle\, \hat{\cal R}_{\vka}\, \hat{\g}^{\vkb}_{m_2n_2}\, 
\hat{\g}^{\vkc}_{m_3n_3}\, \rangle_{k_1} 
&\equiv& \langle\, \hat{\cal R}_{\vka}\, \langle\, \hat{\g}^{\bf k_2}_{m_2n_2}\, 
\hat{\g}^{\vkc}_{m_3n_3}\, \rangle_{k_1}\, \rangle \nn\\
&=& -\,\f{(2\,\pi)^{5/2}}{4\, k_1^3\, k_2^3}\, \f{\nt}{4}\,
\ps(k_1)\, \pt(k_2)\, \Pi^{\bf k_2}_{m_2n_2,m_3n_3}\,
\delta^{3}(\vkb + \vkc),\\
\langle\, \hat{\g}^{\vka}_{m_1n_1}\, \hat{\g}^{\vkb}_{m_2n_2}\, 
\hat{\g}^{\vkc}_{m_3n_3}\, \rangle_{k_3} 
&\equiv& \langle\, \langle\, \hat{\g}^{\vka}_{m_1n_1}\, 
\hat{\g}^{\vkb}_{m_2n_2}\,\rangle_{k_3}\,
\hat{\g}^{\vkc}_{m_3n_3}\,\rangle \nn\\
&=&-\, \f{(2\,\pi)^{5/2}}{4\, k_1^3\, k_3^3}\, \l(\f{\nt-3}{32}\r)\,
\pt(k_1)\,\pt(k_3)\nn\\
& &\times\, \Pi^{\vka}_{m_1n_1,m_2n_2}\, \Pi^{\vkc}_{m_3n_3,ij}\,
{\hat n}_{1i}\,{\hat n}_{1j}\;
\delta^{3}(\vka + \vkb),
\end{eqnarray}
\end{subequations}
where, in the cases of the scalar and the tensor bi-spectra and the 
scalar-scalar-tensor cross-correlation, we have considered $\vkc$ to
be the squeezed mode, while we have considered $\vka$ to be the 
squeezed mode in the case of the scalar-tensor-tensor cross-correlation.
Upon making use of the above expressions for the three-point
functions in the definitions~(\ref{eqs:ngp}) for the non-Gaussianity 
parameters, we can express the consistency relations in the squeezed 
limit as follows: 
\begin{subequations}
\label{eqs:cr}
\begin{eqnarray}
\lim_{k_3\to 0}\, \fnl(\vk,-\vk,\vkc) 
&=& \f{5}{12}\,\l[\ns(k) - 1\r],\label{eq:fnl-cr}\\
\lim_{k_3\to 0}\, \cnls(\vk,-\vk,\vkc) 
&=& \l[\frac{\ns(k) - 4}{4}\r]\, 
\l(\Pi^{\vkc}_{m_3n_3,\bar{m}\bar{n}}\r)^{-1} \,
\Pi^{\vkc}_{m_3n_3,ij}\, {\hat n}_{i}\, {\hat n}_{j},\label{eq:cnls-cr}\\  
\lim_{k_1\to 0}\,\cnlt(\vka,\vk,-\vk)  
&=& \f{\nt(k)}{2}\, \l(\Pi^{\vk}_{m_2n_2,m_3n_3} 
\r)^{-1}\,
\Pi^{\vk}_{m_2n_2,m_3n_3},\label{eq:cnlt-cr}\\
\lim_{k_3\to 0}\, \hnl(\vk,-\vk,\vkc) 
&=& \l[\f{\nt(k) - 3}{2}\r]\, 
\biggl(2\,\Pi^{\vk}_{m_1n_1,m_2n_2}\, \Pi^{\vkc}_{m_3n_3,{\bar m}{\bar n}} 
+ \Pi^{\vk}_{m_1n_1,{\bar m}{\bar n}}\, \Pi^{\vkc}_{m_3n_3,m_2n_2}\nn\\ 
& &+\,\Pi^{\vk}_{{\bar m}{\bar n},m_2n_2}\, \Pi^{\vkc}_{m_3n_3,m_1n_1}\biggr)^{-1}\nn\\
& & \times\,\Pi^{\vk}_{m_1n_1,m_2n_2}\,
\Pi^{\vkc}_{m_3n_3,ij}\, {\hat n}_{i}\, {\hat n}_{j},\label{eq:hnl-cr}
\end{eqnarray}
\end{subequations}
where we have explicitly illustrated the point that $\ns$ and $\nt$ are, 
in general, dependent on the wavenumber [a dependence which can be arrived 
at from the corresponding power spectra through the 
expressions~(\ref{eqs:nsnt})].
It is useful to note here that, during slow roll inflation, while the 
non-Gaussianity parameters $\fnl$ and $\cnlt$ are of the order of the slow 
parameters, the quantities $\cnls$ and $\hnl$ prove to be of order unity.
This does not imply that the parameters $\cnls$ and $\hnl$ are `large'.
They are of order of unity due to the manner in which they have been 
introduced.

\par

Finally, we would like to stress here the fact that we have arrived at the 
above consistency relations essentially assuming that the perturbations 
are initially in the Bunch-Davies vacuum and that the amplitude of the 
scalar and the tensor perturbations are frozen on super-Hubble scales.
While we have focussed on single field models of inflation driven by the 
canonical scalar field, the amplitude of the perturbations are known to
be conserved in any single field model.
For this reason, one can expect the consistency relations to hold even in
non-canonical models of inflation, provided the perturbations are in the
Bunch-Davies vacuum~\cite{kundu-2013}.


\section{Analytical examples}\label{sec:ae}

In this section, we shall explicitly confirm the validity of the above 
consistency relations involving the tensors in two analytically
tractable examples.
We shall first consider a particular case of power law inflation and then
discuss the so-called Starobinsky model which permits a brief period of
departure from slow roll.


\subsection{A power law case}\label{subsec:pl}

Power law inflation corresponds to the situation wherein the scale factor is 
given by
\begin{equation}
a(\eta)=a_1\, \l(\f{\eta}{\eta_1}\r)^{\gamma + 1},
\label{eq:a-p-law}
\end{equation}
where $a_1$ and $\eta_1$ are constants, while $\gamma <-2$. 
In such a situation, the first slow roll parameter proves to be a constant, 
and is given by $\epsilon_1 = (\gamma + 2)/(\gamma +1)$.
Further, since $z\propto a$ in this case, the scalar and the tensor modes 
are described by the same functions. 
One finds that, the solutions to the Mukhanov-Sasaki equations~(\ref{eqs:ms}) 
can be expressed in terms of the Hankel functions of the first kind 
$H_\nu^{(1)}(x)$ as (see, for instance, Refs.~\cite{p-law,hazra-2012})
\begin{equation} 
v_k(\eta) = u_k(\eta)
=\sqrt{\f{-\pi\,\eta}{4}}\; {\rm e}^{-i\,\pi\, (\nu-1/2)/2}\;
H_{-\nu}^{(1)}(-k\,\eta),\label{eq:v-p-law}
\end{equation}
where $\nu = \gamma + 1/2$.
These specific solutions have been arrived at by demanding that 
they satisfy the Bunch-Davies initial conditions at very early 
times (i.e. as $\eta\to -\infty$)~\cite{texts,reviews,bunch-1978}.
The power spectra in power law inflation can be arrived at from the
amplitudes of the Hankel functions, evaluated at late times (i.e. as 
$\eta\to 0$).
One can obtain the scalar and tensor power spectra to be
\begin{equation}
\ps(k)= 16\,\epsilon_1\, \pt(k)
= \f{1}{2\,\pi^3\, \Mpl^2\, \epsilon_1}\,
\l(\f{\vert\eta_1\vert^{\gamma+1}}{a_1}\r)^2\,
\bigl\vert \Gamma[-(\gamma+1/2)]\bigr\vert^2\; 
\l(\f{k}{2}\r)^{2\,(\gamma+2)},
\end{equation}
where, recall that, $\epsilon_1=(\gamma+2)/(\gamma+1)$, and $\Gamma(x)$
represents the Gamma function.
Note that the scalar and tensor spectral indices corresponding to these
power spectra are constants, and are given by $\ns-1 = \nt = 2\,(\gamma+2)$.
If the consistency relations~(\ref{eqs:cr}) are indeed satisfied, then, 
upon setting each of the factors involving the polarization of the tensor 
perturbations to be unity, the above spectral indices would lead to the following 
values of non-Gaussianity parameters of our interest:
\begin{subequations}
\label{eq:cr-pl}
\begin{eqnarray}
\cnls &=& \f{\ns - 4}{4} = \f{2\, \gamma+1}{4},\\
\cnlt &=& \f{\nt}{2} = \gamma+2,\\
\hnl &=& \f{\nt - 3}{8} =  \f{2\, \gamma+1}{8},
\end{eqnarray}
\end{subequations}
which are constants independent of the wavenumber.
Our task now would be to evaluate the three-point functions using the
Maldacena formalism and examine if we indeed arrive at these values
in the squeezed limit.

\par

Ideally, it would have been desirable to arrive at analytic expressions 
describing the three-point functions in power law inflation for an 
arbitrary index $\gamma$. 
This clearly requires having to calculate the various integrals describing
the correlations that we had summarized earlier in Subsec.~\ref{subsec:tpfs-mf}.
In fact, the spectral dependences of the three-point functions in
power law inflation can be easily arrived at (in, say, the equilateral
and the squeezed limits) without actually having to carry out the 
integrals involved~\cite{hazra-2013,sreenath-2013}.
These results for the squeezed limit then immediately point to the fact 
that the non-Gaussianity parameters would be independent of scale. 
But, in order to be able to establish the consistency 
conditions~(\ref{eq:cr-pl}) explicitly, apart from the spectral 
dependences, we shall require the amplitude of the integrals as well.
But, care is required in handling the integrals in the extreme 
sub-Hubble limit (i.e. as $\eta\to -\infty$) wherein the integrands 
oscillate with increasing frequency. 
This aspect seems to make it difficult to carry out the integrals and
express them in a closed analytic form for a generic $\gamma$.

\par

For the above reason, in order to establish the consistency 
relations, we shall focus on the specific case of $\gamma = -3$ or,
equivalently, $\nu = -5/2$. 
In this case, the scalar and tensor modes simplify to
\begin{equation}
f_k(\eta)
= \f{g_k(\eta)}{\sqrt{2}} 
= -\f{1}{\sqrt{2\,k^5}\,\Mp}\, \f{1}{a_1\, \eta_1^2}\,
\l(3+3\,i\,k\,\eta - k^2\,\eta^2\r)\, {\rm e}^{-i\,k\,\eta},
\end{equation}
so that the corresponding derivatives are given by
\begin{equation}
f_k'(\eta)
= \f{g_k'(\eta)}{\sqrt{2}} 
= \f{-i}{\sqrt{2\,k^3}\,\Mp}\,\f{1}{a_1\, \eta_1^2}\,
\l(k^2\,\eta^2 - i\,k\,\eta\r)\, {\rm e}^{-i\,k\,\eta}.
\end{equation}
As $\eta \to 0$, the scalar and tensor modes reduce to
\begin{equation}
\lim_{\eta\to 0} f_k(\eta)
= \lim_{\eta\to 0} \f{g_k(\eta)}{\sqrt{2}} 
= -\f{3}{\sqrt{2\,k^5}\,\Mp}\, \f{1}{a_1\, \eta_1^2}.
\end{equation}
We can arrive at the three-point functions of interest upon substituting
the above scalar and tensor modes, their derivatives and their asymptotic 
behavior at late times, in the expressions~(\ref{eq:Grrg}), (\ref{eq:Grgg})
and~(\ref{eq:Gggg}), and evaluating the various integrals involved.
We find that, upon setting the factors containing the polarization tensor
to be unity, in the squeezed limit, the three-point functions of interest 
are given by 
\begin{subequations}
\begin{eqnarray}
\lim_{k_3\to 0}\, k^3\,k_3^3\, G_{\cR\cR\g}(\vk,-\vk,\vkc) 
&=& \f{5}{4}\; \l(\f{3}{\Mpl\, a_1\, \eta_1^2}\right)^4\, 
\f{1}{k^2\,k_3^2},\\
\lim_{k_1\to 0}\, k_1^3\,k^3\, G_{\cR\g\g}(\vka,\vk,-\vk) 
&=&  \l(\f{3}{\Mpl\, a_1\, \eta_1^2}\right)^4\, 
\f{1}{k_1^2\,k^2},\\  
\lim_{k_3\to 0}\, k^3\,k_3^3\, G_{\g\g\g}(\vk,\vk,-\vkc) 
&=& \f{5}{2}\,\l(\f{3}{\Mpl\, a_1\, \eta_1^2}\right)^4\, 
\f{1}{k^2\,k_3^2}.
\end{eqnarray}
\end{subequations}
The non-Gaussianity parameters corresponding to these three-point
functions can be easily obtained to be $\cnls = -5/4$, $\cnlt = -1$ 
and $\hnl = -5/8$. 
These values exactly match the results~(\ref{eq:cr-pl}) with
$\gamma = -3$, which implies that the consistency conditions are
indeed satisfied in this case.


\subsection{The case of the Starobinsky model}\label{subsec:sm}
 
The second example that we shall consider is the Starobinsky model. 
In the Starobinsky model, the inflaton rolls down a linear potential 
which changes its slope suddenly at a particular value of the scalar 
field~\cite{starobinsky-1992}. 
The governing potential is given by
\begin{equation} \label{starobinsky potential}
V(\phi)
=\left\{ \begin{array}{rcl}                  
V_0 + A_+(\phi -\phi_0) &{\rm for}& \phi > \phi_0,\\  
V_0 + A_-(\phi -\phi_0) &{\rm for}& \phi < \phi_0,
\end{array}\right.
\end{equation}
where $V_0$, $A_+$, $A_-$ and $\phi_0$ are constants. 
As should be clear, there is a discontinuity in the slope of the potential 
at $\phi_0$. 
This discontinuity leads to a brief period of deviation from slow 
roll as the field traverses $\phi_0$, before slow roll is 
restored~\cite{starobinsky-1992,martin-2012,arroja-2011-2012,martin-2014}.

\par

In order to calculate the three-point functions of our interest, evidently, 
we shall require the behavior of the scale factor, the first slow roll 
parameter $\epsilon_1$, and the scalar and the tensor modes 
[cf.~Eqs.~(\ref{eq:Grrg}), (\ref{eq:Grgg}) and~(\ref{eq:Gggg})].
An important aspect of the Starobinsky model---which permits the background,
the perturbations and the different correlation functions to be calculated 
analytically---is the assumption that the constant $V_0$ is the dominant term 
in the potential that is driving the field as it crosses the discontinuity 
at~$\phi_0$.
In such a situation, we can consider the Hubble parameter to be a constant,
say, $H_0$, determined by the relation $H_0^2\simeq V_0/(3\, \Mp^2)$.
Due to this reason, the scale factor essentially corresponds to that of de 
Sitter and, hence, the first slow parameter $\epsilon_1$ remains small 
throughout the evolution. 
In fact, under the assumption that $V_0$ is suitably large compared to the
other terms in the potential, the first slow roll parameter before and after 
the field crosses $\phi_0$ can be obtained to 
be~\cite{starobinsky-1992,martin-2012,arroja-2011-2012,martin-2014} 
\begin{subequations}
\begin{eqnarray}\label{eq:srp1}
\epsilon_{1+}(\eta) 
& \simeq & \f{A_+^2}{18\,\Mp^2\,H_0^4},\label{srpm}\\ 
\epsilon_{1-}(\eta) 
& \simeq & \f{A_-^2}{18\,\Mp^2\,H_0^4}\,
\l[1-\f{\Delta A}{A_-}\,\l(\f{\eta}{\eta_0}\r)^3\r]^2,\label{eq:e1m}
\end{eqnarray}
\end{subequations}
where $\Delta A=A_--A_+$, and $\eta_0$ denotes the conformal time when the 
transition takes place.
Note that, in the above expressions and some that follow, a plus or minus 
sign in the sub-script (or, when convenient, in the super-script) denote 
the quantities before and after the transition at $\eta_0$.

\par

It should be clear from the above expressions for the first slow roll 
parameter that its value changes as the field traverses across $\phi_0$.
Recall that, the second slow roll parameter is defined as $\epsilon_2
={\dot \epsilon}_1/(H\,\epsilon_1)$. 
Because of the change in the value of $\epsilon_1$ at $\eta_0$, the 
magnitude of the second slow roll parameter $\epsilon_2$ exhibits a 
sharp rise leading to brief period of departure from slow roll, 
before slow roll is restored again at a suitably later time.    
In order to calculate the non-vanishing contributions to the three-point 
functions in the squeezed limit of our interest, it suffices for us to
know the scalar modes, and we do not require its time derivative 
which involves $\epsilon_2$~\cite{martin-2012,martin-2014}. 
This, in turn, implies that we do not need the actual behavior of the 
the second slow roll parameter.
Interestingly, due to the simple nature of the potential and the assumptions 
that one works under, one finds that the scalar modes are given by the 
standard Bunch-Davies modes in the de Sitter limit both before and after
the transition.
However, because of the transition that occurs at $\eta_0$, the modes 
post-transition are related to the modes prior to the transition by the 
standard Bogoliubov coefficients.
One finds that the modes before and after the transition are given 
by~\cite{starobinsky-1992,martin-2012,arroja-2011-2012,martin-2014}:
\begin{subequations}
\begin{eqnarray}
f_k^{+}(\eta)
&=&\frac{i\, H_0}{2\, \Mp\, \sqrt{{k^3}\,\epsilon_{1+}}}\,
\l(1+i\,k\,\eta\right)\,{\rm e}^{-i\,k\,\eta},\label{eq:fk-bt}\\
f_k^{-}(\eta)
&=&\frac{i\,H_0\,\alpha_k}{2\,\Mp\,\sqrt{{k^3}\,\epsilon_{1-}}}\,
\l(1+i\,k\,\eta\r)\, {\rm e}^{-i\,k\,\eta}
-\frac{i\,H_0\,\beta_k}{2\,\Mp\,\sqrt{{k^3}\,\epsilon_{1-}}}
\l(1-i\,k\,\eta\right)\, {\rm e}^{i\,k\,\eta},\label{eq:fk-at}
\end{eqnarray}
\end{subequations}
respectively.
The quantities $\alpha_k$ and $\beta_k$ are the Bogoliubov coefficients,
which can be determined by matching the modes at the transition.
The Bogoliubov coefficients are found to be
\begin{subequations}
\begin{eqnarray}
\alpha_k 
&=& 1+\frac{3\,i\,\Delta A}{2\,A_{+}}\;\frac{k_0}{k}\,
\left(1+\frac{k_0^2}{k^2}\right),
\label{eq:alphak-sm}\\
\beta_k 
&=& -\frac{3\,i\,\Delta A}{2\,A_+}\;\f{k_0}{k}\,
\l(1+\frac{i\, k_0}{k}\r)^2\, {\rm e}^{2\,i\,k/k_{0}},
\label{eq:betak-sm}
\end{eqnarray}
\end{subequations}
with $k_0 = -1/\eta_0 = a_0\,H_0$, and $a_0$ being the value of the
scale factor at the transition.

\par 

In contrast to the scalar modes, the evolution of the tensor modes are
determined only by the behaviour of the scale factor.
Since the scale factor always remains that of de Sitter, the tensor modes
are not affected by the transition at $\phi_0$, and are given by the standard 
Bunch-Davies solution, viz.
\begin{equation}
g_k(\eta) = \f{i\,\sqrt{2}\;H_0}{\Mpl\,\sqrt{2\,k^3}}\,
\l(1+i\,k\,\eta\r)\, {\rm e}^{-i\,k\,\eta},\label{eq:gk-ds}
\end{equation}
over the complete domain in time that is of interest to us.
The time derivative of the tensor mode, which we shall require, is given by
\begin{equation}
g_k'(\eta) = \f{i\,\sqrt{2}\;H_0}{\Mpl\,\sqrt{2\,k^3}}\,
k^2\,\eta\, {\rm e}^{-i\,k\,\eta}.\label{eq:gkp-ds}
\end{equation}

\par

The power spectra can be easily arrived at from the above scalar and
tensor modes.
At late times, i.e. as $\eta\to 0$, the scalar power spectrum can be
obtained to be
\begin{eqnarray}
\ps(k) 
&=& \l(\f{H_0}{2\,\pi}\r)^2\,
\l(\f{3\,H_0^2}{A_-}\r)^2\, 
\vert \alpha_k - \beta_k \vert^2 \nn\\
&=& \l(\f{H_0}{2\,\pi}\r)^2\, 
\l(\f{3\,H_0^2}{A_-}\r)^2\, 
\l[\cI(k) + \cI_{\rm c}(k)\, \cos\l(\f{2\,k}{k_0}\r) 
+ \cI_{\rm s}(k)\, \sin\l(\f{2\,k}{k_0}\r)\r],\label{eq:sps-sm}
\end{eqnarray}
where the quantities $\cI(k)$, $\cI_{\rm c}(k)$ and $\cI_{\rm s}(k)$ are 
given by
\begin{subequations}
\begin{eqnarray}
\cI(k) 
&=& 1 + \f{9}{2}\, \l(\f{\Delta A}{A_+}\r)^2\, \l(\f{k_0}{k}\r)^2 
+ 9\, \l(\f{\Delta A}{A_+}\r)^2\, \l(\f{k_0}{k}\r)^4 
+ \f{9}{2}\, \l(\f{\Delta A}{A_+}\r)^2\, \l(\f{k_0}{k}\r)^6,\\
\cI_{\rm c}(k) 
&=& \f{3\,\Delta A}{2\,A_+}\, \l(\f{k_0}{k}\r)^2\, 
\l[\l(\f{3\,A_-}{A_+} - 7\r) 
- \f{3\,\Delta A}{A_+}\, \l(\f{k_0}{k}\r)^4\r],\\
\cI_{\rm s}(k) 
&=& -\f{3\, \Delta A}{A_+}\, \f{k_0}{k}\, 
\l[1 + \l(\f{3\,A_-}{A_+} - 4 \r)\, \l(\f{k_0}{k}\r)^2 
+ \f{3\,\Delta A}{A_+}\, \l(\f{k_0}{k}\r)^4 \right].
\end{eqnarray}
\end{subequations}
The above scalar power spectrum exhibits a step-like feature with two nearly 
scale invariant rungs (that correspond to the two domains of slow roll) and 
a burst of oscillations (associated with the modes that leave the Hubble radius 
during the period of fast roll) connecting the two rungs (in this context, see, 
for instance, Fig.~3 of Ref.~\cite{martin-2012}).
Since the tensor modes are given by the standard Bunch-Davies solution, 
the resulting tensor spectrum is strictly scale invariant, with the 
amplitude being given by
\begin{equation}
\pt(k) = \f{2\,H_0^2}{\pi^2\,\Mp^2}.\label{eq:tps-sm}
\end{equation}
As in the power law case discussed in the previous sub-section, in order
to establish the consistency relations, we first need to evaluate the 
scalar and tensor spectral indices from the above power spectra.
We then need to obtain the non-Gaussianity parameters $\cnls$, $\cnlt$ and 
$\hnl$ using the consistency conditions~(\ref{eqs:cr}), and compare them
with the non-Gaussianity parameters evaluated from the squeezed limit of
the corresponding three-point functions. 

\par

The scalar spectral index $\ns$ corresponding to the above scalar power 
spectrum can be obtained to be 
\begin{eqnarray}
\ns(k) &=& \f{1}{2}\;
\biggl[\cI(k) + \cI_{\rm c}(k)\, \cos\l(\f{2\,k}{k_0}\r) 
+ \cI_{\rm s}(k)\, \sin\l(\f{2\,k}{k_0}\r)\biggr]^{-1}\;
\biggl\{8\,\cI(k) - 3\, \cJ(k)\nn\\ 
& &+ \l[8\,\cI_{\rm c}(k) - 3\, \cJ_{\rm c}(k)\r]\, \cos \l(\f{2\,k}{k_0} \r) 
+ \l[8\,\cI_{\rm s}(k) - 3\, \cJ_{\rm s}(k)\r]\, 
\sin \l(\f{2\,k}{k_0} \r)\biggr\},
\end{eqnarray}
where $\cJ(k)$, $\cJ_{\rm c}(k)$ and $\cJ_{\rm s}(k)$ are given by
\begin{subequations}
\label{eqs:cJ}
\begin{eqnarray}
\cJ(k) 
&=& 2 + 15\,\l(\f{\Delta A}{A_+}\r)^2\, \l(\f{k_0}{k}\r)^2 
+ 42\, \l(\f{\Delta A}{A_+}\r)^2\, \l(\f{k_0}{k}\r)^4 
+ 27\, \l(\f{\Delta A}{A_+}\r)^2\, \l(\f{k_0}{k}\r)^6, \\
\cJ_{\rm c}(k) 
&=& \f{\Delta A}{A_+}\, 
\l[4 + 3\, \l(\f{9\,A_-}{A_+} - 17\r)\, \l(\f{k_0}{k} \r)^2 
+ \f{12\,\Delta A}{A_+}\, \l(\f{k_0}{k}\r)^4 
- \f{27\,\Delta A}{A_+}\, \l(\f{k_0}{k}\r)^6\r],\qquad\quad\\
\cJ_{\rm s}(k) 
&=& \f{2\, \Delta A}{A_+}\, \f{k_0}{k}\, 
\l[\l(\f{3\,A_-}{A_+} -11\r) 
- 6\, \l(\f{3\,A_-}{A_+}-4\r)\, \l(\f{k_0}{k}\r)^2
- \f{27\,\Delta A}{A_+}\, \l(\f{k_0}{k}\r)^4 \right].
\end{eqnarray}
\end{subequations}
From this expression for $\ns$, upon suitably ignoring overall factors containing
the polarization tensor, we can obtain the non-Gaussianity parameter $\cnls$ 
to be 
\begin{eqnarray}
\cnls(k) 
= \f{\ns(k) - 4}{4}
= -\,\f{3}{8}\; \f{\cJ(k) + \cJ_{\rm c}(k)\, \cos \l(2\,k/k_0\r) 
+ \cJ_{\rm s}(k)\, \sin\l(2\,k/k_0\r)}{\cI(k) 
+ \cI_{\rm c}(k)\, \cos \l(2\,k/k_0\r) 
+ \cI_{\rm s}(k)\, \sin \l(2\,k/k_0\r)}.\label{eq:cnls-sm-cr}
\end{eqnarray}
We shall require the tensor spectral index $\nt$ in order to evaluate 
the other two non-Gaussianity parameters $\cnlt$ and $\hnl$ using the
consistency relations~(\ref{eqs:cr}).
However, since the tensor power spectrum~(\ref{eq:tps-sm}) is strictly 
scale invariant at the level of approximation we are working with, the 
corresponding spectral index $\nt$ vanishes identically.
Moreover, note that since the tensor modes remain unaffected by the 
transition, the tensor bi-spectrum will be of the same form as in the
de Sitter case, a situation wherein it is easy to establish analytically
that $\hnl=-3/8$ in the squeezed limit (see, for instance, 
Refs.~\cite{maldacena-2003,tensor-bs,sreenath-2013}).
In order to establish the consistency relation for the parameter $\cnlt$, 
we shall evaluate the tensor spectral index numerically, and compare the 
result with the analytical expressions that we shall obtain from the
Maldacena formalism for the three-point functions in the squeezed limit.

\par

\begin{figure}[!htb]
\begin{center}
\includegraphics[width=15.0cm]{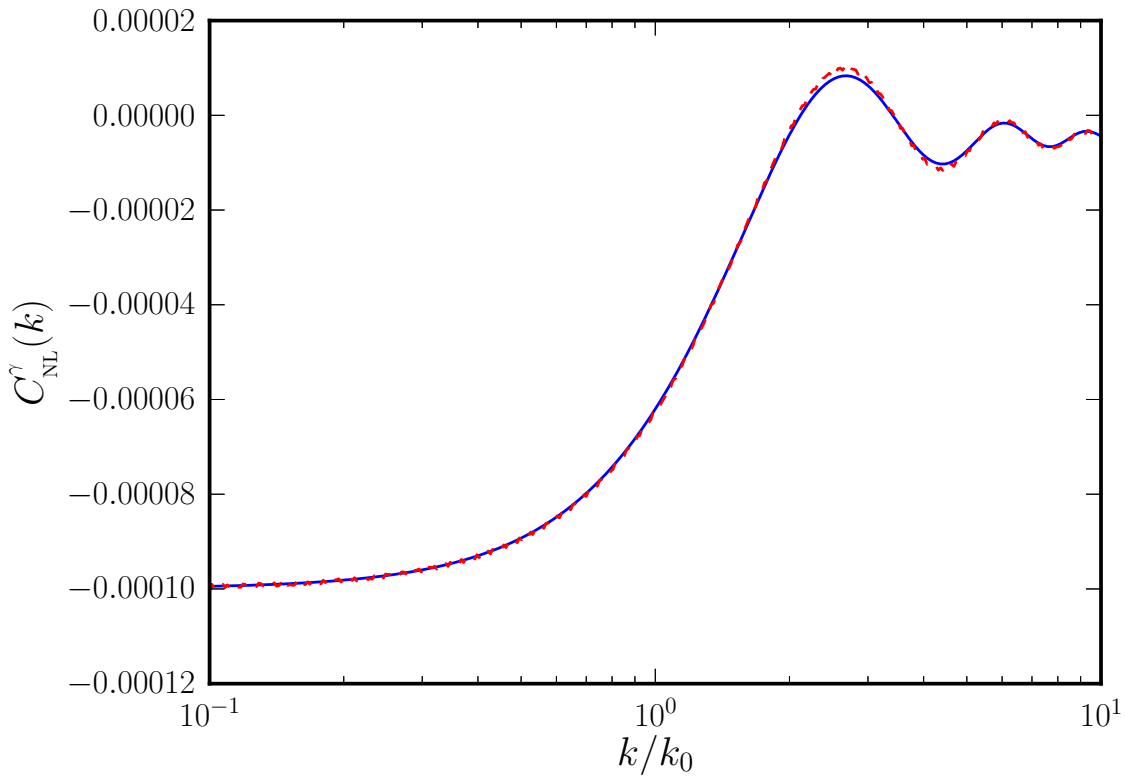} 
\caption{\label{fig:sm-atc-anr}
The non-Gaussianity parameter $\cnlt$ in the Starobinsky model, evaluated 
in the squeezed limit, has been plotted as a function of $k/k_0$. 
The solid blue curve represents the parameter arrived at from the analytical 
results for the scalar-tensor-tensor cross-correlation (obtained using the
Maldacena formalism) and the scalar and the tensor power spectra.
The dashed red curve corresponds to the non-Gaussianity parameter obtained
from consistency condition~(\ref{eq:cnlt-cr}), with the tensor spectral index
being determined numerically.
Evidently, there is good agreement between the two results, indicating that
the consistency relation holds even when departures from slow roll occur.
Note that we have worked with the following values of the potential parameters
in arriving at these results: $\phi_0/\Mpl=0.707$, $V_0/\Mpl^4=2.37\times 
10^{-12}$, $A_+/\Mpl^3=3.35\times 10^{-14}$ and $A_-/\Mpl^3=7.26\times10^{-15}$.
These values have been chosen so that the assumptions of the Starobinsky model,
under which the analytical results have been arrived at, are valid (in this 
context, see, for instance, Ref.~\cite{martin-2012}).}
\end{center}
\end{figure}

The scalar-scalar-tensor cross-correlation in the Starobinsky model can be 
calculated analytically by dividing the integrals involved into two parts, 
corresponding to the epochs before and after the transition, and making use 
of the above expressions for the first slow roll parameter and the scalar 
and the tensor modes.
In the squeezed limit of the tensor mode, on ignoring the polarization
tensors, we find that the scalar-scalar-tensor three-point function can 
be written as
\begin{equation}
\lim_{k_3\to 0}\, k^3\, k_3^3\; 
G_{\sr\sr\gamma}^{m_3n_3}(\vk,-\vk,\vkc) 
= \f{27\,H_0^8}{8\,\Mp^2\, A_-^2}\, 
\l[\cJ(k) + \cJ_{\rm c}(k)\, \cos\l(\f{2\,k}{k_0}\r) 
+ \cJ_{\rm s}(k)\, \sin \l(\f{2\,k}{k_0}\r)\r],
\end{equation}
with $\cJ(k)$, $\cJ_{\rm c}(k)$ and $\cJ_{\rm s}(k)$ being given by
Eqs.~(\ref{eqs:cJ}).
Upon making use of this expression and the power spectra~(\ref{eq:sps-sm})
and~(\ref{eq:tps-sm}) in the definition~(\ref{eq:cnls}) of the parameter
$\cnls$ (and suitably ignoring the factors involving the polarization tensors), 
we find that one exactly arrives at the result~(\ref{eq:cnls-sm-cr}), thereby
establishing the consistency relation for this case. 
Similarly, in the squeezed limit of the scalar mode, scalar-tensor-tensor 
correlation can be obtained to be
\begin{equation}
\lim_{k_1\to 0}\, k_1^3\, k^3\,
G_{\sr\gamma\gamma}^{m_2n_2m_3n_3}(\vka,\vk,-\vk) 
= \f{H_0^4}{8\, \Mp^4}\, 
\l[\cK (k) + \cK_{\rm c}(k)\, \cos\l(\f{2\,k}{k_0}\r) 
+ \cK_{\rm s}(k)\, \sin\l(\f{2\,k}{k_0}\r)\r],
\end{equation}
where the quantities $\cK(k)$, $\cK_{\rm c}(k)$ and $\cK_{\rm s}(k)$ are 
given by
\begin{subequations}
\begin{eqnarray}
\cK(k) 
&=& 4\,\l(\f{A_-}{A_+}\r)^2 
+ 9\, \l(\f{\Delta A}{A_+}\r)^2\, \l(\f{k_0}{k}\r)^6,\\
\cK_{\rm c}(k) 
&=& \f{3\, \Delta A}{A_+}\, \l(\f{k_0}{k}\r)^2\,
\l[2 + 6\, \f{\Delta A}{A_+}\, \l(\f{k_0}{k}\r)^2
- \f{3\, \Delta A}{A_+}\, \l(\f{k_0}{k}\r)^4\r],\\
\cK_{\rm s}(k) 
&=& \f{3\, \Delta A}{A_+}\, \l(\f{k_0}{k}\r)^3\, 
\l[\l(\f{3\,A_-}{A_+} - 4 \r) 
- \f{6\, \Delta A}{A_+}\, \l(\f{k_0}{k}\r)^2\r].
\end{eqnarray}
\end{subequations}
In Fig.~\ref{fig:sm-atc-anr}, we have plotted the $\cnlt$ that results from 
the above analytical expression for the scalar-tensor-tensor cross-correlation 
and the power spectra~(\ref{eq:sps-sm}) and~(\ref{eq:tps-sm}).
In the same figure, we have also plotted the $\cnlt$ that arises from the 
numerical determination of the tensor spectral index and the consistency 
condition~(\ref{eq:cnlt-cr}).     
It is clear from the figure that these two quantities match very well,
indicating the fact that the consistency relation is valid in this case
as well.

 
\section{Numerical investigation of scenarios involving deviations 
from slow roll}\label{sec:ne}

In the last section, we had investigated the validity of the consistency 
relations comprising of the tensor perturbations in two specific situations 
that had proved to be analytically tractable. 
It is well known that the scalar consistency relation is valid in slow 
roll inflation~\cite{maldacena-2003,creminelli-2004,cr-rd}, and its 
applicability in situations containing departures from slow roll has 
also been confirmed in a few instances (for analytical examples, see 
Refs.~\cite{cr-d-d-sr-ar} and, for some numerical results, see
Refs.~\cite{cr-d-d-sr-nr}).
Our general arguments in Sec.~\ref{sec:cr-sl} as well as the analysis of 
the Starobinsky model in the previous section suggest that the consistency 
relations involving tensors too can be expected to be valid even in 
scenarios consisting of deviations from slow roll.  
It will be interesting to explicitly examine these relations in different
models containing brief periods of fast roll.

\par 

While a nearly scale invariant inflationary scalar power spectrum is quite
consistent with the CMB and other cosmological data~\cite{be-fim,martin-2013-14}, 
it has been repeatedly noticed that certain features in the primordial scalar 
power spectrum fit the data better (for recent discussions, see, for instance, 
Refs.~\cite{rc}).
Typically, one finds that one or more of the following features lead to an
improved fit to the data: (i)~a sharp drop in power on large scales (see
Refs.~\cite{pi}; for some recent discussions in this context, see
Refs.~\cite{lp-ls}), (ii)~a short burst of oscillations over an intermediate 
range of scales~\cite{l-22-40,hazra-2010,benetti-2011,benetti-2013,ci}, and
(iii)~small and persisting oscillations that extend over a wide range of
scales~\cite{pso,pahud-2009,flauger-2010,kobayashi-2011,aich-2013,meerburg-2013,easther-2013}.
In fact, it is exactly these three type of possibilities that have been 
considered by the Planck team in their analysis of probable features in 
the scalar power spectrum~\cite{planck-2013-ci}.
In this section, we shall consider three inflationary models that lead to 
features of the different types mentioned above and investigate numerically
whether the consistency relations of our interest remain valid in these 
non-trivial situations as well.

\par

The three inflationary models that we shall consider are as follows.
The first example that we shall focus on is the inflationary model 
governed by the potential
\begin{equation}
V(\phi) = \f{m^2}{2}\,\phi^2 
-\f{\sqrt{2\,\lambda\,(n-1)}\, m}{n}\, \phi^n
+\f{\lambda}{4}\,\phi^{2\,(n-1)},\label{eq:p-pi}
\end{equation}
which contains a point of inflection at 
\begin{equation}
\phi_0=\l[\f{2\, m^2}{(n-1)\, \lambda}\r]^{\f{1}{2\, (n-2)}}.
\end{equation}
For certain values of the parameters, this model permits two epochs of 
slow roll inflation, with a brief period of departure from inflation 
sandwiched in between, a scenario that has been dubbed punctuated 
inflation~\cite{pi}. 
The sudden deviation from inflation and the rapid return to slow roll 
results in a steep drop in the scalar power on large scales, which can
help fit the lower power seen at the large angular scales in the CMB.
Notably, these models predict a rise in the tensor power over scales
wherein the scalar power drop sharply.
The second model that we shall consider is the quadratic potential 
containing a step that has been introduced by hand.   
The complete potential is given by the 
expression~\cite{l-22-40,hazra-2010,benetti-2011}
\begin{equation}
V(\phi) = \f{m^2}{2}\,\phi^2\,
\l[1 + \alpha\,\tanh\,\l(\f{\phi -\phi_0}{\Delta\phi}\r)\r],
\label{eq:p-qp-ws}
\end{equation}
where, clearly, $\alpha$ and $\Delta \phi$ denote the height and the width 
of the step, while $\phi_0$ represents its location.
The step in the potential leads to a brief period of fast roll which, in
turn, leads to a burst of oscillations in the power spectrum.
These oscillations have been shown to improve the fit to the CMB data around 
the multipoles of $\ell=20$--$50$.
The last case that we shall consider is a potential motivated by string 
theory known as axion monodromy model.
The model is described by the following 
potential~\cite{flauger-2010,aich-2013,easther-2013}:
\begin{equation}
V(\phi) = \lambda\, \l[\phi 
+ \alpha\, \cos\,\l(\f{\phi}{\beta} +\delta\r)\r].\label{eq:p-amm}
\end{equation}
The continued oscillations in the potential lead to persistent oscillations
in the inflationary scalar power spectrum. 
(For an illustration of the scalar power spectra that arise in these three 
models, we would refer the reader to Fig.~9 of Ref.~\cite{hazra-2013}.)

\par

We shall now numerically examine the validity of the consistency relations 
involving the tensor perturbations in the above-mentioned models.
As we had mentioned, in an earlier work, we had developed a numerical procedure 
and had constructed a Fortran code for evaluating the three-point scalar-tensor 
cross-correlations and the tensor bi-spectrum~\cite{sreenath-2013}.
We shall make use of the code to compute the three-point functions in the 
squeezed limit as well as the scalar and the tensor spectral indices, in 
order to check the consistency conditions~(\ref{eqs:cr}). 
We shall work with values of the parameters for the potentials that have 
been shown to lead to an improved fit to the CMB data (in this context,
see Refs.~\cite{sreenath-2013,hazra-2013}). 
While evaluating the scalar and tensor power spectra, it has long been
known that, in order to arrive at spectra with good accuracy, it is 
sufficient to numerically integrate the modes from a time when they are
well inside the Hubble radius [say, when $k/(a\, H)\simeq 10^2$] to a time when 
they are suitably outside [say, when $k/(a\, H)\simeq 10^{-5}$]~\cite{ne-ps}.   
It has been shown that, while evaluating the three-point functions in the 
Maldacena formalism, it suffices to carry out the numerical calculations 
{\it roughly}\/ over a similar time domain~\cite{sreenath-2013,hazra-2013,ng-ne,ng-f}. 
However, there are three points that we need to emphasize in this regard. 
Firstly, to achieve higher levels of numerical accuracy, say, of the order
of $1$--$3\%$ or better, for the three-point functions, one may have to integrate 
from a time when the modes are deeper inside the Hubble radius than $k/(a\, H)
\simeq 10^2$.
In our calculations, we shall choose to integrate from $k/(a\, H)\simeq 10^3$
for the punctuated inflation model~(\ref{eq:p-pi}) and the quadratic potential 
with a step~(\ref{eq:p-qp-ws}).
The oscillating nature of the potential~(\ref{eq:p-amm}) in the axion monodromy 
model leads to certain resonant behavior (see the first of the references in
Ref.~\cite{cr-d-d-sr-ar} and Refs.~\cite{hazra-2013,sreenath-2013}), and it 
typically requires one to integrate from further deep inside the Hubble radius,
even in the case of the power spectrum. 
For this reason, we shall choose to integrate from  $k/(a\, H)\simeq 10^4$ in
this case.
Secondly, due to the rapid oscillations of the modes when they are inside
the Hubble radius, a cut off in the integrands is required to regulate the
integrals at early times\footnote{In theory, such a cut-off is mandatory to 
identify the correct perturbative vacuum~\cite{maldacena-2003,ng-ncsf,martin-2012}.}.
We shall work with a cut off of the form $\exp-[\kappa\, k/(a\, H)]$, where 
$\kappa$ is a parameter that has to be chosen according to the initial time
from which the integrations are to be carried out.
For instance, the earlier the initial time, the smaller the quantity $\kappa$ 
has to be~\cite{sreenath-2013,hazra-2013,ng-ne,ng-f}.
Since, we shall integrate from $k/(a\, H)\simeq 10^3$ in the cases of punctuated 
inflation and the quadratic potential with a step, we shall work with $\kappa=1/50$ 
in these cases, which is known to lead to a good 
accuracy~\cite{hazra-2013,sreenath-2013}.
However, as we integrate from deeper inside the Hubble radius in the axion 
monodromy model, we shall work with $\kappa = 1/500$ in this case.
The third and the last point concerns the implementation of the squeezed limit.
To achieve this limit, we shall work with the smallest wavenumber (say, the 
largest scale mode that leaves the Hubble radius at the earliest possible time) 
that is numerically tenable.
As a result, inherently, there will arise a weak wavenumber dependent effect 
when attempting to establish the consistency conditions numerically, with the 
small scale modes satisfying the condition better than the longer ones.

\par

\begin{figure}[!t]
\begin{center}
\begin{tabular}{ccc}
\hskip -10pt
\includegraphics[width=5.25cm]{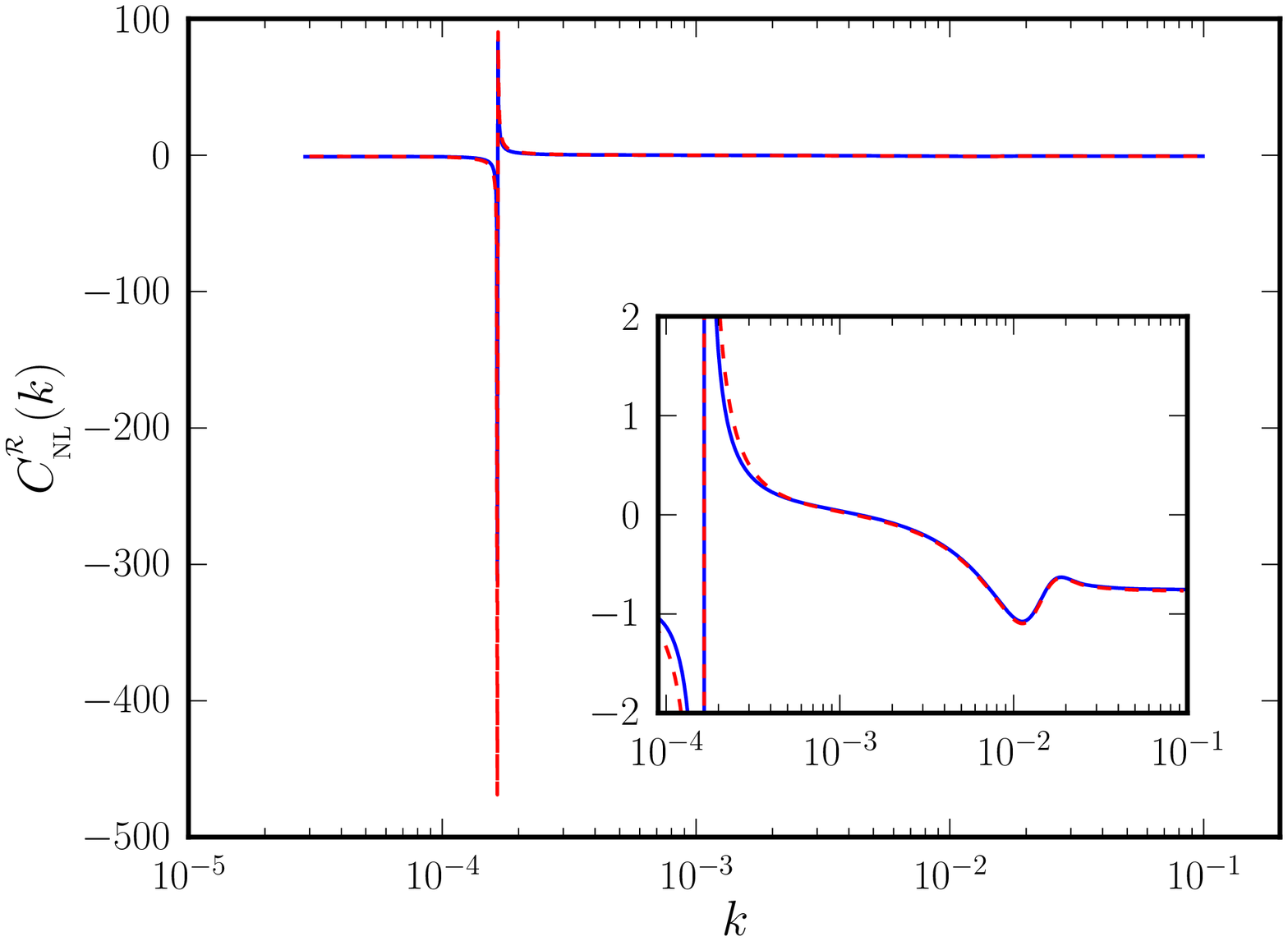} & 
\hskip -10pt
\includegraphics[width=5.25cm]{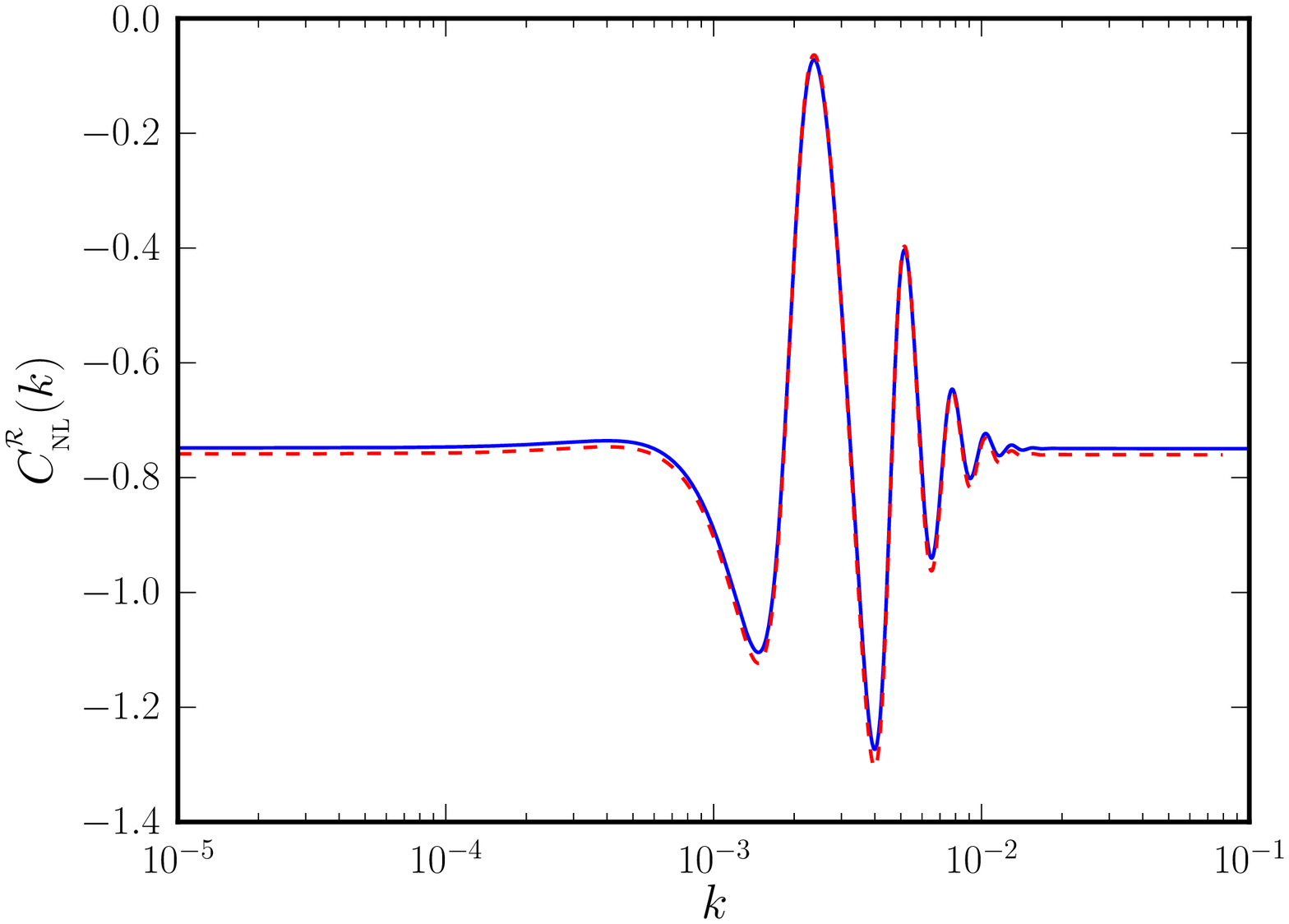} &
\hskip -10pt
\includegraphics[width=5.25cm]{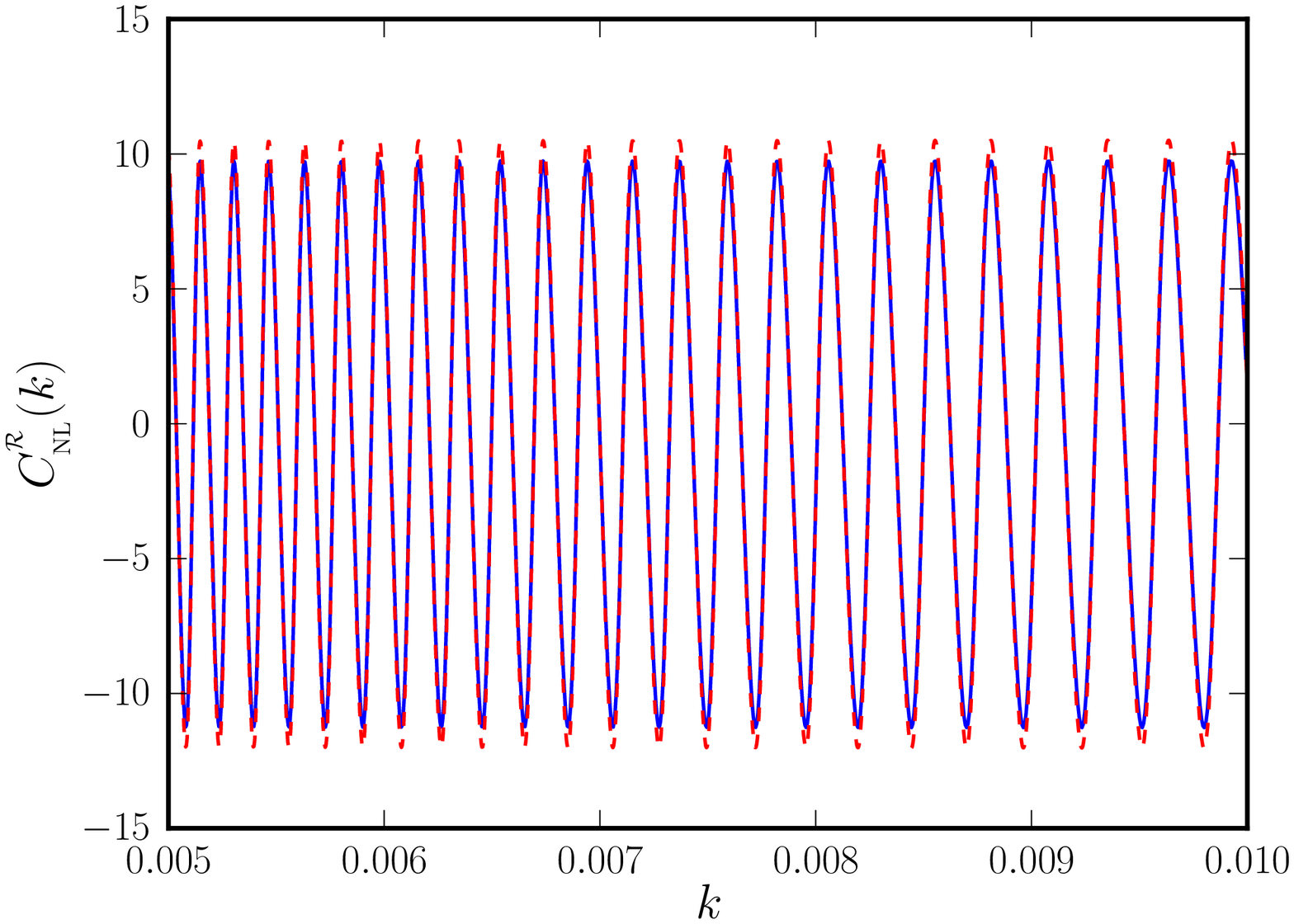} \\
\hskip -10pt
\includegraphics[width=5.25cm]{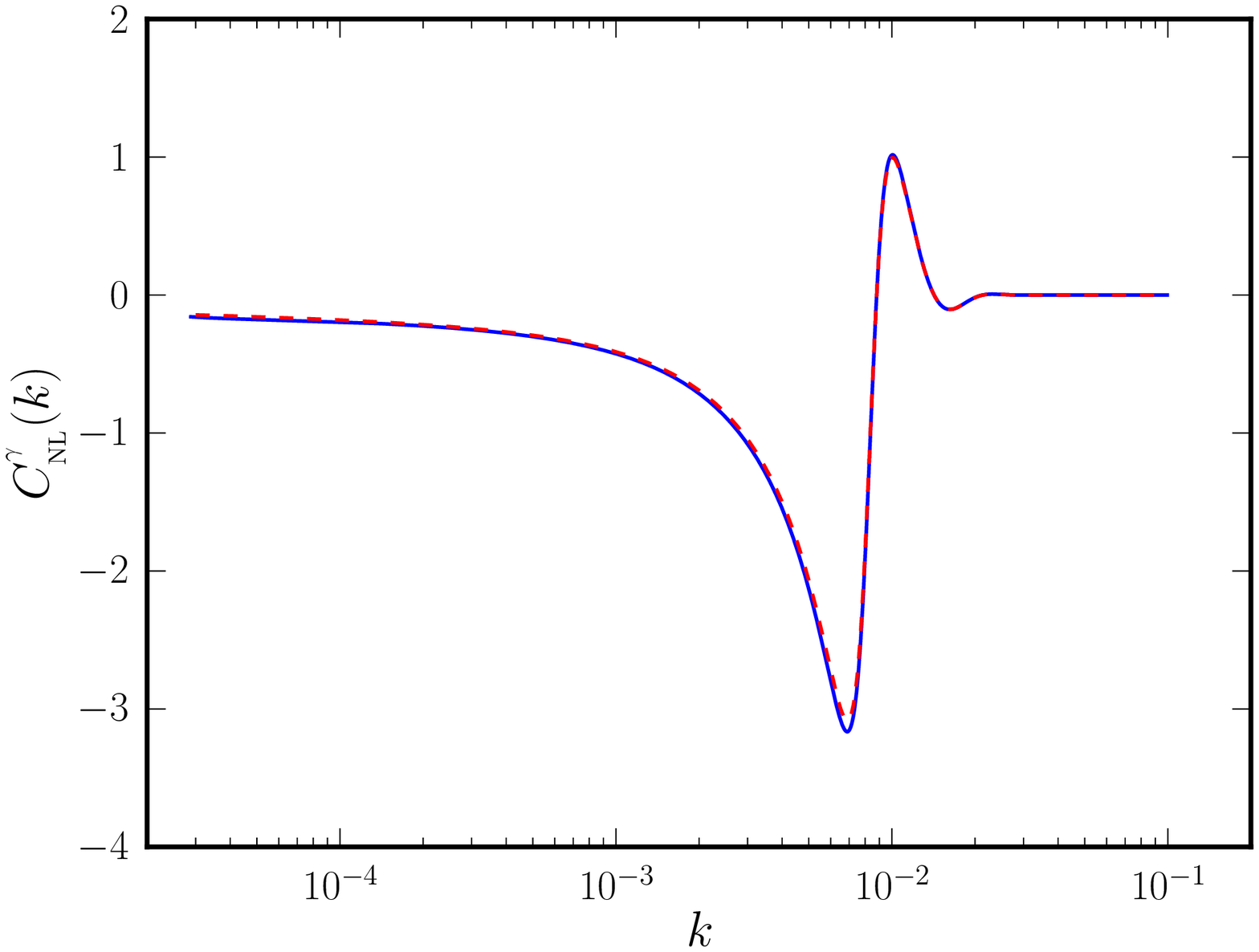} &
\hskip -10pt
\includegraphics[width=5.25cm]{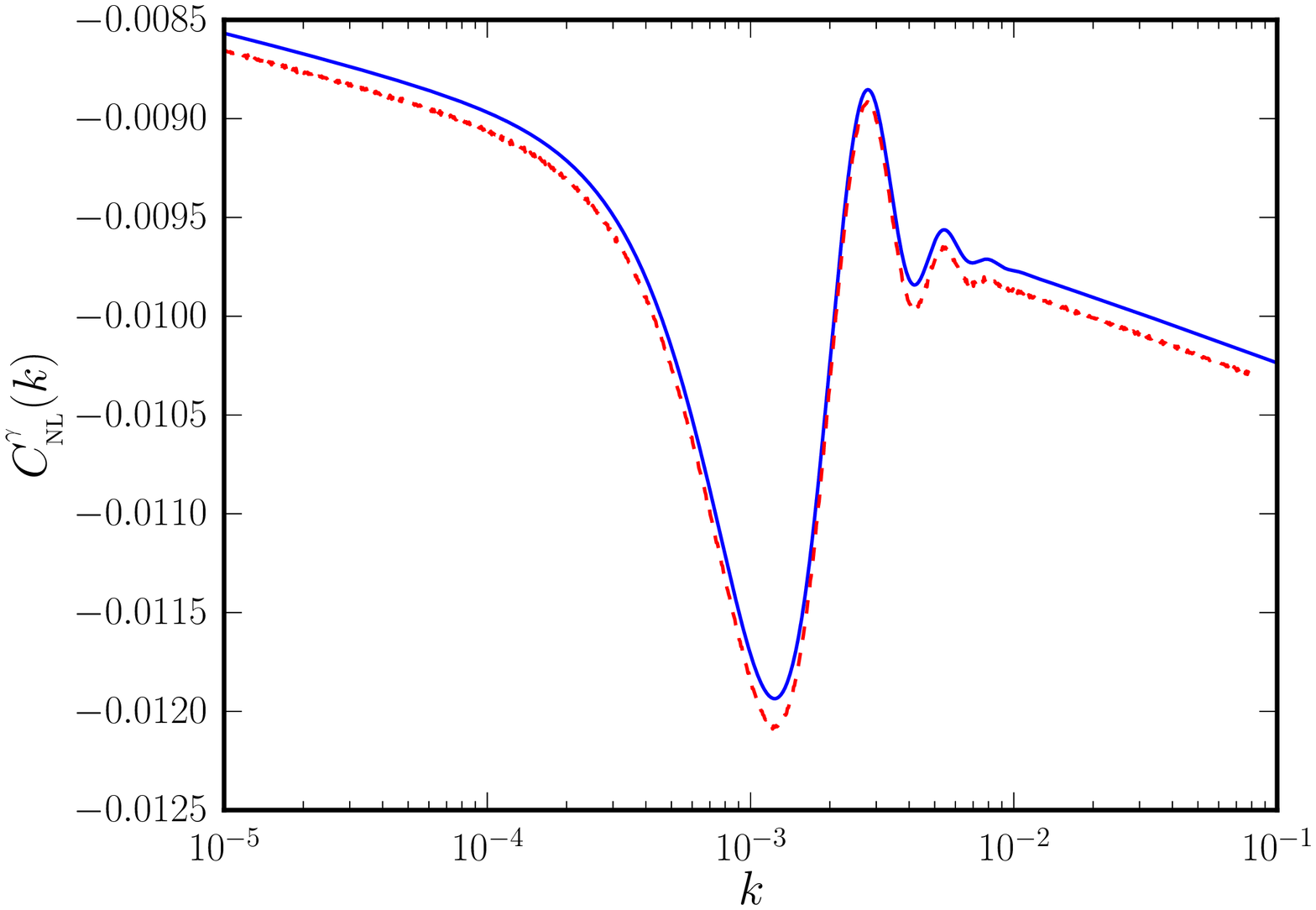} &
\hskip -10pt
\includegraphics[width=5.25cm]{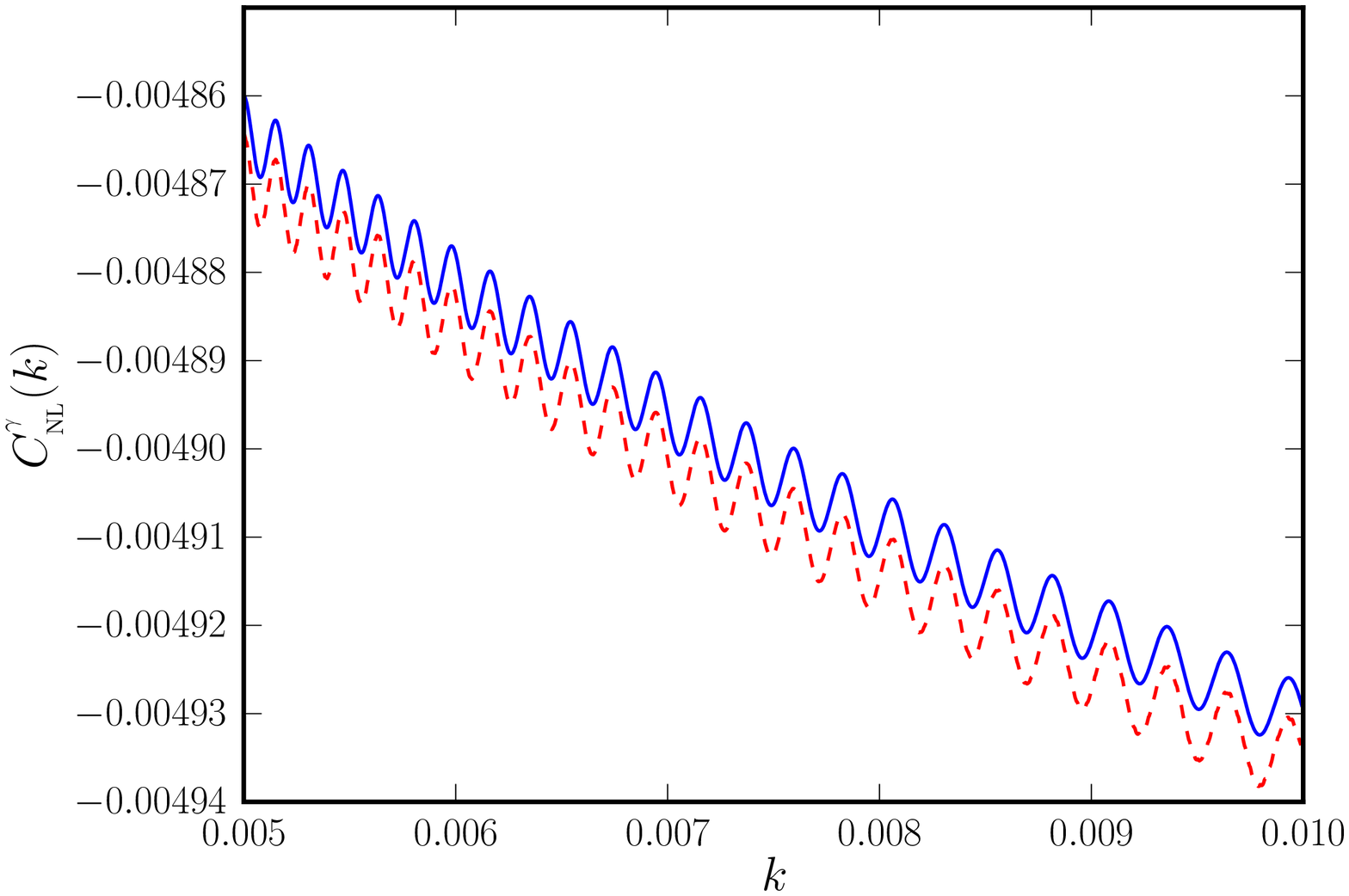} \\
\hskip -10pt
\includegraphics[width=5.25cm]{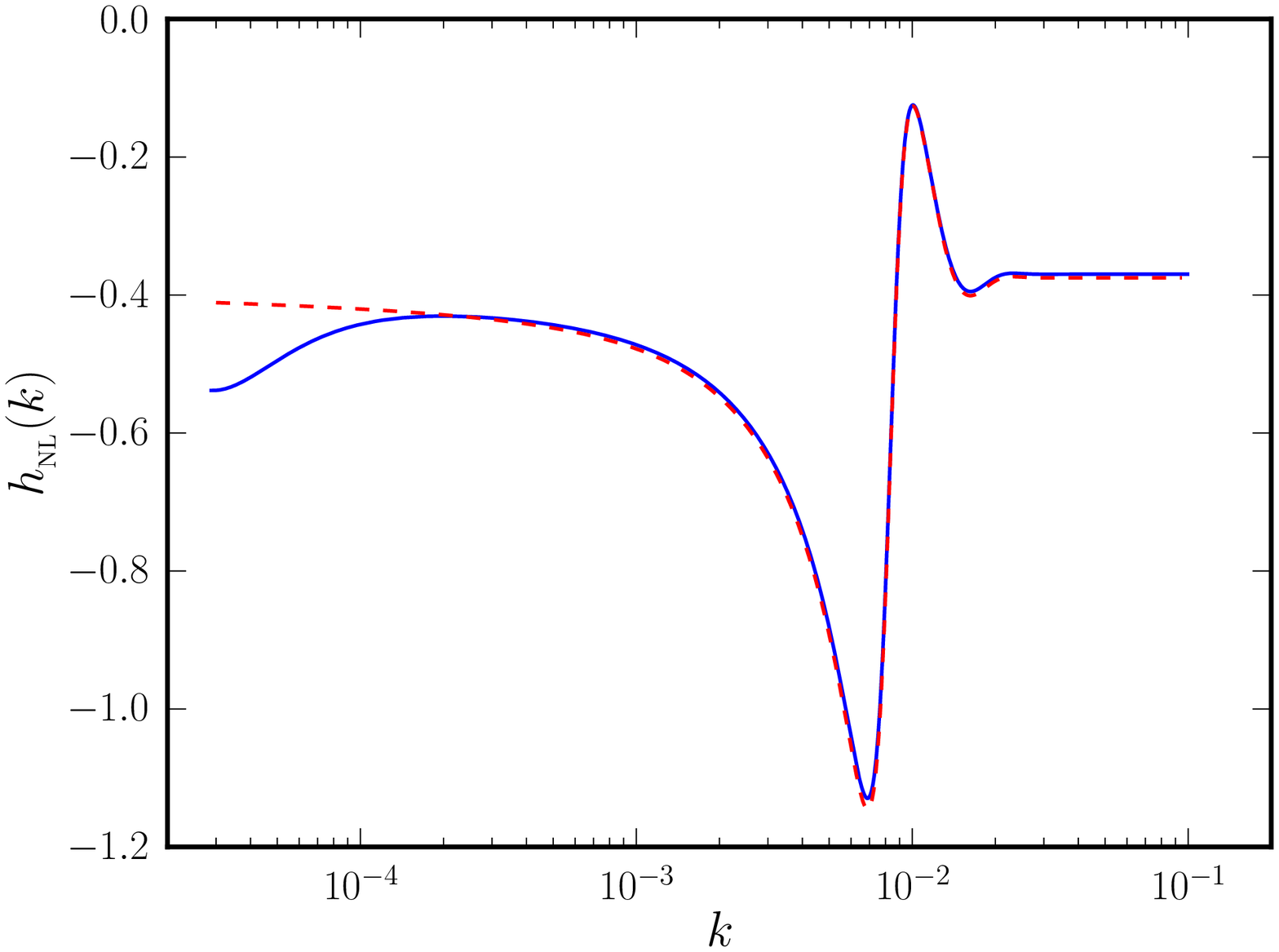} &
\hskip -10pt
\includegraphics[width=5.25cm]{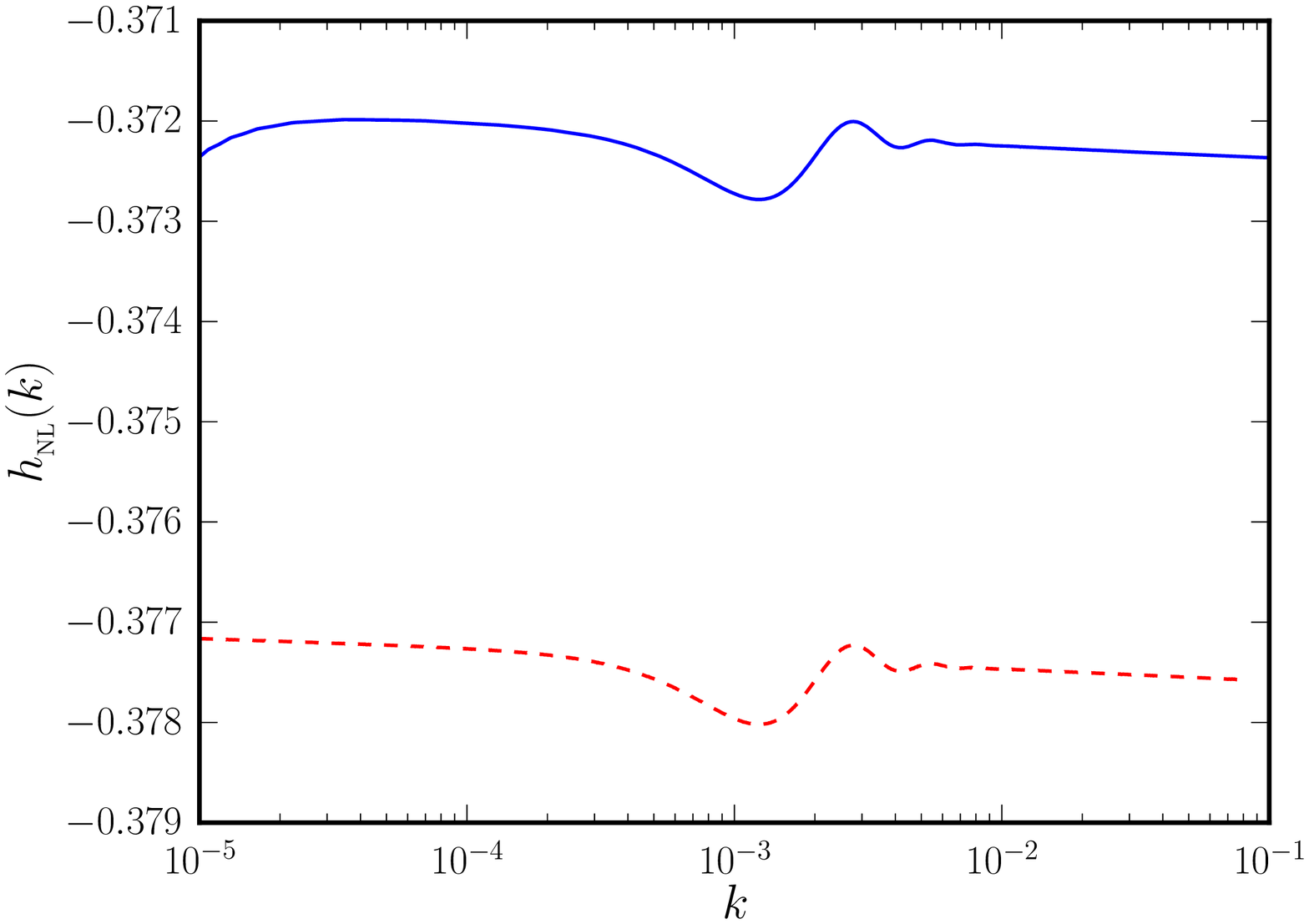} &
\hskip -10pt
\includegraphics[width=5.25cm]{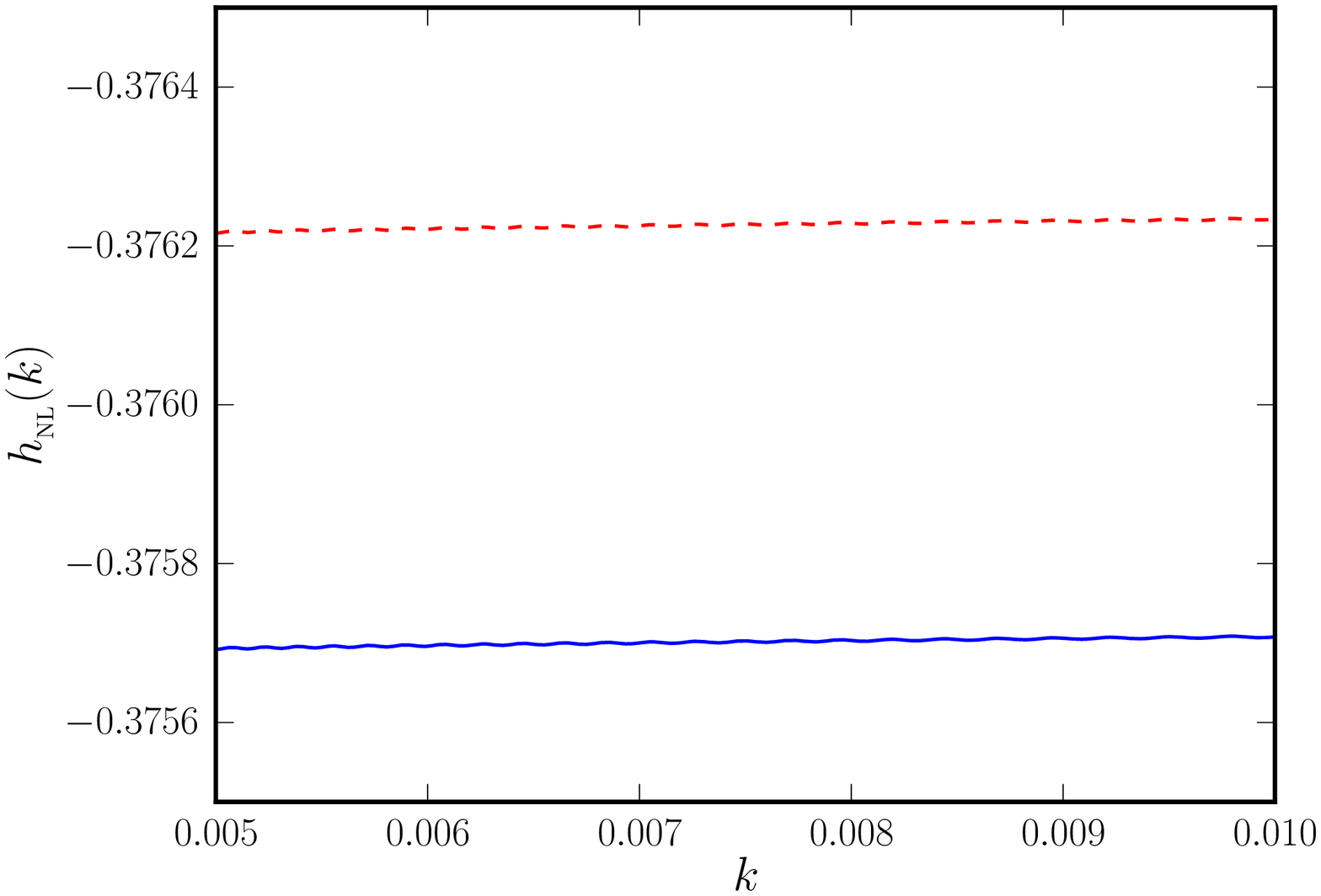} 
\end{tabular}
\caption{\label{fig:atc-pi-qpws-amm} 
The non-Gaussianity parameters,  $\cnls$ (the top row), $\cnlt$ (the middle 
row) and $\hnl$ (the bottom row), arrived at using the Maldacena formalism 
and from the scalar and tensor spectral indices through the consistency conditions,
have been plotted as a function of the wavenumber for the three inflationary 
models of our interest, viz. punctuated inflation (the left column), the 
quadratic potential with a step (the middle column) and the axion monodromy 
model (the right column). 
While the solid blue lines correspond to the numerical results for the 
parameters obtained using the Maldacena formalism, the dashed red lines 
represent the values arrived at from the spectral indices and the 
consistency relations. 
The two results match at the level of $3\%$ or better in the cases of 
punctuated inflation and the quadratic potential with the step, with 
the largest differences arising for the smallest wavenumbers due to the 
inherent limitation in implementing the squeezed limit numerically.
Though the match is slightly poorer in the axion monodromy model (with 
differences of the order of $7\%$ for certain wavenumbers), we find that 
the match can be improved by integrating for a longer duration. 
At a first look, the difference in the cases of the model with the step 
and the axion monodromy model in the last row may seem striking.
But, we should clarify here that it is simply due to the fact that the 
non-Gaussianity parameter $\hnl$ has been plotted in these cases over 
a rather small range in amplitude to highlight the mild variations.
Needless to add, these figures confirm the validity of the consistency 
relations in single field inflationary models even in situations that
allow strong departures from slow roll.}
\end{center}
\end{figure}

In Fig.~\ref{fig:atc-pi-qpws-amm}, we have plotted the numerical results for 
the non-Gaussianity parameters $\cnls$, $\cnlt$ and $\hnl$ for the three models 
discussed above. 
We find that the results from the spectral indices match the numerical results 
for the non-Gaussianity parameters obtained using Maldacena formalism at 
the level of $3\%$ or better in the cases of punctuated inflation and the
quadratic potential with the step, with the largest differences arising for 
the smallest wavenumbers for the reasons discussed above. 
The match is slightly poorer in the axion monodromy model 
(with differences of about $7\%$ for some wavenumbers), 
but the match improves if we carry out the integrals over a longer duration 
in time.  
(We should mention here that the seeming difference in the cases of the 
model with the step and the axion monodromy model in the last row of the
figure is due to the fact that the non-Gaussianity parameter $\hnl$ has 
been plotted over a rather small range in amplitude to highlight the
mild variations.)
Clearly, the consistency relations hold true even in inflationary models 
that contain deviations from slow roll inflation. 


\section{Discussion}\label{sec:d}

Consistency relations link the three-point functions in the squeezed limit 
to the scalar and tensor power spectra.
They essentially indicate that, in the squeezed limit, the three-point functions
carry the same extent of information as the two-point functions do.
Evidently, the consistency conditions can be conveniently expressed as relations 
between the non-Gaussianity parameters (two of which we had recently 
introduced~\cite{sreenath-2013}) and the scalar or the tensor spectral indices. 
The consistency relations arise essentially because of the fact that inflaton 
is the only clock in single field inflationary models.
Due to this reason, the amplitude of the long wavelength modes freeze in such 
situations.
As we had discussed in Sec.~\ref{sec:cr-sl}, this implies that, in the squeezed 
limit of interest, the long wavelength modes simply modify the background 
spatial coordinates. 
In this work, we have explicitly examined, both analytically and numerically, 
the validity of consistency relations involving the tensor perturbations in
single field inflationary models.
Corroborating the general arguments that have been outlined earlier, we find
that the consistency conditions hold true even in non-trivial scenarios 
involving drastic deviations from slow roll, such as it occurs in the case 
of punctuated inflation.

\par 

It should be evident that the consistency conditions provide a remarkable 
tool to observationally confirm if inflation was driven by a single scalar 
field. 
Clearly, any observed deviation from the conditions would point to other
scenarios such as inflation driven by multiple scalar fields or even exotic 
possibilities such as the ekpyrotic scenario.
It would be worthwhile to repeat some of our analyses for non-canonical scalar 
fields, and then go to study the extent of deviations from the consistency 
conditions that arise in inflationary models comprising of more than one 
scalar field.
We are currently investigating some of these issues.


\section*{Acknowledgements}

The authors wish to thank J\'er\^ome Martin for discussions as well as 
detailed comments on the manuscript. 
They would also like to thank Dhiraj Hazra and Rajeev Jain for discussions. 
We acknowledge the use of the high performance computing facility at the
Indian Institute of Technology Madras, Chennai, India.



\begin{thebibliography}{99}
\bibitem{texts}
E.~W.~Kolb and M.~S.~Turner, {\sl The Early Universe}\/ (Addison-Wesley, 
Redwood City, California, 1990); S.~Dodelson, {\sl Modern Cosmology}\/ 
(Academic Press, San Diego, U.S.A., 2003); V.~F.~Mukhanov, {\sl Physical 
Foundations of Cosmology}\/ (Cambridge University Press, Cambridge, 
England, 2005); S.~Weinberg, {\sl Cosmology}\/ (Oxford University Press, 
Oxford, England, 2008); R.~Durrer, {\sl The Cosmic Microwave Background}\/ 
(Cambridge University Press, Cambridge, England, 2008); D.~H.~Lyth and 
A.~R.~Liddle, {\sl The Primordial Density Perturbation}\/ (Cambridge 
University Press, Cambridge, England, 2009); P.~Peter and J-P.~Uzan, 
{\sl Primordial Cosmology}\/ (Oxford University Press, Oxford, England, 
2009); H.~Mo, F.~v.~d.~Bosch and S.~White, {\sl Galaxy Formation and 
Evolution}\/ (Cambridge University Press, Cambridge, England, 2010);
D.~Baumann and L.~McAllister, {\sl Inflation and String Theory},\/
arXiv:1404.2601 [hep-th].
\bibitem{reviews} 
H.~Kodama and M.~Sasaki, Prog.\ Theor.\ Phys.\ Suppl.\ {\bf 78}, 1 (1984); 
V.~F.~Mukhanov, H.~A.~Feldman and R.~H.~Brandenberger, Phys.\ Rep.\ {\bf 215}, 
203 (1992); J.~E.~Lidsey, A.~Liddle, E.~W.~Kolb, E.~J.~Copeland, T.~Barreiro 
and M.~Abney, Rev.\ Mod.\ Phys.\ {\bf 69}, 373 (1997); 
A.~Riotto, arXiv:hep-ph/0210162; W.~H.~Kinney, astro-ph/0301448; J.~Martin, 
Lect. Notes Phys. {\bf 738}, 193 (2008); J.~Martin, Lect. Notes Phys. {\bf 669}, 
199 (2005); J.~Martin, Braz. J. Phys. {\bf 34}, 1307 (2004); B.~Bassett, 
S.~Tsujikawa and D.~Wands, Rev.\ Mod.\ Phys.\ {\bf 78}, 537 (2006); 
W.~H.~Kinney, arXiv:0902.1529 [astro-ph.CO]; L.~Sriramkumar, Curr.\ Sci.\ 
{\bf 97}, 868 (2009) [arXiv:0904.4584 [astro-ph.CO]]; 
D.~Baumann, arXiv:0907.5424v1 [hep-th].
\bibitem{wmap-2011}
D.~Larson {\it et al.},\/ Astrophys.\ J.\ Suppl.\ {\bf 192}, 16 (2011);
E.~Komatsu {\it et al.},\/ Astrophys.\ J.\ Suppl.\ {\bf 192}, 18 (2011).
\bibitem{wmap-2013}
C.~L.~Bennett {\it et al.},\/ Astrophys.\ J.\ Suppl.\ {\bf 208}, 20 (2013); 
G.~Hinshaw {\it et al.},\/ Astrophys.\ J.\ Suppl.\ {\bf 208}, 19 (2013).
\bibitem{planck-2013-cmbps}
P.~A.~R.~Ade {\it et al.},\/ arXiv:1303.5075 [astro-ph.CO].
\bibitem{planck-2013-ccp}
P.~A.~R.~Ade {\it et al.},\/ arXiv:1303.5076 [astro-ph.CO].
\bibitem{planck-2013-ci}
P.~A.~R.~Ade {\it et al.},\/ arXiv:1303.5082 [astro-ph.CO].
\bibitem{be-fim}
J.~Martin, C.~Ringeval and R.~Trotta, Phys.\ Rev.\ D\ {\bf 83}, 063524
(2011); M.~J. Mortonson, H.~V.~Peiris and R.~Easther, Phys.\ Rev.\ D\ 
{\bf 83}, 043505 (2011); R.~Easther and H.~Peiris, Phys.\ Rev.\ D\ {\bf 85}, 
103533 (2012); J.~Norena, C.~Wagner, L.~Verde, H.~V.~Peiris and 
R.~Easther, Phys.\ Rev.\ D\ {\bf 86}, 023505 (2012).
\bibitem{martin-2013-14}
J.~Martin, C.~Ringeval and V.~Vennin, arXiv:1303.3787 [astro-ph.CO];
J.~Martin, C.~Ringeval, R.~Trotta and V.~Vennin, JCAP {\bf 1403}, 039 (2014);
arXiv:1405.7272 [astro-ph.CO].
\bibitem{earlyng} 
A.~Gangui, F.~Lucchin, S.~Matarese and S.~Mollerach, Astrophys.\ J.\ 
{\bf 430}, 447 (1994); A.~Gangui, Phys.\ Rev.\ D\ {\bf 50}, 
3684 (1994); A.~Gangui and J.~Martin, Mon.\ Not.\ Roy.\ Astron.\ Soc.\ 
{\bf 313}, 323 (2000); L.~Wang and M.~Kamionkowski, Phys.\ Rev.\ D\
{\bf 61}, 063504 (2000).
\bibitem{maldacena-2003}
J.~Maldacena, JHEP\ {\bf 0305}, 013 (2003).
\bibitem{ng-ncsf}
D.~Seery and J.~E.~Lidsey, JCAP {\bf 0506}, 003 (2005); X.~Chen, Phys.\ 
Rev.\ D {\bf 72}, 123518 (2005); X.~Chen, M.-x.~Huang, S.~Kachru and 
G.~Shiu, JCAP {\bf 0701}, 002 (2007); D.~Langlois, S.~Renaux-Petel, 
D.~A.~Steer and T.~Tanaka, Phys.\ Rev.\ Lett.\ {\bf 101}, 061301 (2008); 
Phys.\ Rev.\ D\ {\bf 78}, 063523 (2008).
\bibitem{ng-reviews}
X.~Chen, Adv.\ Astron.\ {\bf 2010}, 638979 (2010); Y.~Wang, 
arXiv:1303.1523 [hep-th].
\bibitem{ng-da}
E.~Komatsu and D.~N.~Spergel, Phys.\ Rev.\ D\ {\bf 63}, 063002 (2001);
E.~Komatsu, D.~N.~Spergel and B.~D.~Wandelt, Astrophys.\ J.\ {\bf 634},
14 (2005); D.~Babich and M.~Zaldarriaga, Phys.\ Rev.\ D\ {\bf 70}, 083005 
(2004); M.~Liguori, F.~K.~Hansen, E.~Komatsu, S.~Matarrese and A.~Riotto, 
Phys.\ Rev.\ D {\bf 73}, 043505 (2006); C.~Hikage, E.~Komatsu and 
T.~Matsubara, Astrophys.\ J.\ {\bf 653} (2006) 11 (2006); J.~R.~Fergusson 
and E.~P.~S. Shellard, Phys.\ Rev.\ D\ {\bf 76}, 083523 (2007); 
A.~P.~S.~Yadav, E.~Komatsu and B.~D.~Wandelt, Astrophys.\ J.\ {\bf 664}, 
680 (2007); P.~Creminelli, L.~Senatore and M.~Zaldarriaga, JCAP {\bf 0703},
019 (2007); A.~P.~S.~Yadav and B.~D.~Wandelt, Phys.\ Rev.\ Lett.\ {\bf 100}, 
181301 (2008); C.~Hikage, T.~Matsubara, P.~Coles, M.~Liguori, F.~K.~Hansen 
and S.~Matarrese, Mon.\ Not.\ Roy.\ Astron.\ Soc.\ {\bf 389}, 1439 (2008);
O.~Rudjord, F.~K.~Hansen, X.~Lan, M.~Liguori, D.~Marinucci and S.~Matarrese,
Astrophys.\ J.\ {\bf 701}, 369 (2009); K.~M.~Smith, L.~Senatore and 
M.~Zaldarriaga, JCAP {\bf 0909}, 006 (2009); J.~Smidt, A.~Amblard, 
C.~T.~Byrnes, A.~Cooray, A.~Heavens and D.~Munshi, Phys.\ Rev.\ D\ {\bf 81}, 
123007 (2010); J.~R.~Fergusson, M.~Liguori and E.~P.~S.~Shellard, 
arXiv:1006.1642v2 [astro-ph.CO].
\bibitem{ng-da-reviews}
M.~Liguori, E.~Sefusatti, J.~R.~Fergusson and E.~P.~S.~Shellard, Adv.\ 
Astron.\ {\bf 2010}, 980523 (2010); A.~P.~S.~Yadav and B.~D.~Wandelt, 
arXiv:1006.0275v3 [astro-ph.CO]; E.~Komatsu, Class.\ Quantum Grav.\
{\bf 27}, 124010 (2010).
\bibitem{planck-2013-cpng}
P.~A.~R.~Ade {\it et al.}, arXiv:1303.5084 [astro-ph.CO].
\bibitem{tensor-bs}
J.~Maldacena and G.~L.~Pimentel, JHEP {\bf 1109}, 045 (2011); X.~Gao, 
T.~Kobayashi, M.~Yamaguchi and J.~Yokoyama, Phys.\ Rev.\ Lett.\ {\bf 107}, 
211301 (2011).
\bibitem{tanaka-2011}
T.~Tanaka and Y.~Urakawa, JCAP {\bf 1105}, 014 (2011).
\bibitem{cc}
X.~Gao, T.~Kobayashi, M.~Shiraishi, M.~Yamaguchi, J.~Yokoyama and
S.~Yokoyama, arXiv:1207.0588 [astro-ph.CO].
\bibitem{cc2}
D.~Jeong and M.~Kamionkowski, Phys.\ Rev.\ Lett.\ {\bf 108}, 251301 (2012);
L.~Dai, D.~Jeong and M.~Kamionkowski, Phys.\ Rev.\ D\ {\bf 87}, 103006 (2013);
Phys.\ Rev.\ D\ {\bf 88}, 043507 (2013). 
\bibitem{sreenath-2013}
V.~Sreenath, R.~Tibrewala and L.~Sriramkumar, JCAP {\bf 1312}, 037 (2013).
\bibitem{kundu-2013}
S.~Kundu, arXiv:1311.1575 [astro-ph.CO].
\bibitem{bicep2}
P.~A.~R.~Ade {\it et. al.},\/ arXiv:1403.3985 [astro-ph.CO];
P.~A.~R.~Ade {\it et. al.},\/ arXiv:1403.4302 [astro-ph.CO].
\bibitem{creminelli-2004}
P.~Creminelli and M.~Zaldarriaga, JCAP {\bf 0410}, 006 (2004).
\bibitem{cr-rd}
C.~Cheung, A.~L.~Flitzpatrick, J.~Kaplan and L.~Senatore, JCAP {\bf 0802},
021 (2008); S.~Renaux-Petel, JCAP {\bf 1010}, 020 (2010); J.~Ganc and 
E.~Komatsu,  JCAP {\bf 1012}, 009 (2010); P.~Creminelli, G.~D'Amico, 
M.~Musso and J.~Norena, JCAP {\bf 1111}, 038 (2011); D.~Chialva, JCAP 
{\bf 1210}, 037 (2012); K.~Schalm, G.~Shiu and T.~van der Aalst, JCAP 
{\bf 1303}, 005 (2013); E.~Pajer, F.~Schmidt and M.~Zaldarriaga, Phys.\
Rev.\ D\ {\bf 88}, 083502 (2013).
\bibitem{npfs}
L.~Senatore and M.~Zaldarriaga, JCAP {\bf 1208}, 001 (2012); P.~Creminelli, 
J.~Norena and M.~Simonovoc, JCAP {\bf 1207}, 052 (2012); P.~Creminelli, 
A.~Perko, L.~Senatore, M.~Simonovic and G.~Trevisan, JCAP {\bf 1311}, 015 (2013);
L.~Berezhiani and J.~Khoury, JCAP {\bf 1402}, 003 (2014); L.~Berezhiani, 
J.~Khoury and J.~Wang, arXiv:1401.7991 [hep-th]; H.~Collins, R.~Holman and 
T.~Vardanyan, arXiv:1405.0017 [hep-th].
\bibitem{e-dfsr}
S.~M.~Leach and A.~R.~Liddle, Phys.\ Rev.\ D\ {\bf 63}, 043508 (2001);
S.~M.~Leach, M.~Sasaki, D.~Wands and A.~R.~Liddle, {\it ibid.} {\bf 64}, 
023512 (2001); R.~K.~Jain, P.~Chingangbam and L.~Sriramkumar, JCAP 
{\bf 0710}, 003 (2007).
\bibitem{p-law}
L.~F.~Abbott and M.~B.~Wise, Nucl.\ Phys.\ B\ {\bf 244}, 541 (1984); 
D.~H.~Lyth and E.~D.~Stewart, Phys.\ Lett.\ B\ {\bf 274}, 168 (1992); 
J.~Martin and D.~J.~Schwarz, Phys.\ Rev.\ D\ {\bf 57}, 3302 (1998); 
L.~Sriramkumar and T.~Padmanabhan, Phys. Rev. D 71, 103512 (2005).
\bibitem{hazra-2012}
D.~K.~Hazra, J.~Martin and L.~Sriramkumar, Phys.\ Rev.\ D\ {\bf 86}, 
063523 (2012). 
\bibitem{bunch-1978}
T.~Bunch and P.~C.~W.~Davies, Proc.\ Roy.\ Soc.\ Lond.\ A\ {\bf 360}, 117 
(1978).
\bibitem{hazra-2013}
D.~K.~Hazra, L.~Sriramkumar and J.~Martin, JCAP {\bf 05}, 026 (2013).
\bibitem{starobinsky-1992}
A.~A.~Starobinsky, Sov.\ Phys.\ JETP\ Lett.\ {\bf 55}, 489 (1992).
\bibitem{martin-2012}
J.~Martin and L.~Sriramkumar, JCAP {\bf 1201}, 008 (2012).
\bibitem{arroja-2011-2012}
F.~Arroja, A.~E.~Romano and M.~Sasaki, Phys.\ Rev.\ D\ {\bf 84}, 123503
(2011); F.~Arroja and M.~Sasaki, JCAP {\bf 1208}, 012 (2012).
\bibitem{martin-2014}
J.~Martin, L.~Sriramkumar and D.~K.~Hazra, arXiv:1404.6093 [astro-ph.CO].
\bibitem{cr-d-d-sr-ar}
J.~Martin, H.~Motohashi and T.~Suyama, Phys.\ Rev.\ D\ {\bf 87}, 023514 (2013); 
M.~H.~Namjoo, H.~Firouzjahi and M.~Sasaki, Europhys.\ Lett.\ {\bf 101}, 39001 (2013); 
M.~G.~Jackson and G.~Shiu, Phys.\ Rev.\ D\ {\bf 88}, 123511 (2013)
\bibitem{cr-d-d-sr-nr}
P.~Adshead, W.~Hu, C.~Dvorkin and H.~V.~Peiris, Phys.\ Rev.\ D\ {\bf 84}, 
043519 (2011); A.~Achucarro, J-O.~Gong, G.~A.~Palma and S.~P.~Patil, Phys.\ 
Rev.\ D {\bf 87}, 121301 (2013); J.~Gong, K.~Schalm and G.~Shiu, Phys.\ Rev.\ 
D\ {\bf 89}, 063540 (2014).
\bibitem{rc}
D.~K.~Hazra, A.~Shafieloo and T.~Souradeep, JCAP {\bf 1307}, 031 (2013);
P.~Hunt and S.~Sarkar, arXiv:1308.2317 [astro-ph.CO].
\bibitem{pi}
R.~K.~Jain, P.~Chingangbam, J.-O.~Gong, L.~Sriramkumar and T.~Souradeep, 
JCAP {\bf 0901}, 009 (2009); R.~K.~Jain, P.~Chingangbam, L.~Sriramkumar 
and T.~Souradeep, Phys.\ Rev.\ D\ {\bf 82}, 023509 (2010). 
\bibitem{lp-ls}
L.~Lello, D.~Boyanovsky and R.~Holman, arXiv:1307.4066 [astro-ph.CO];
M.~Cicoli, S.~Downes and B.~Dutta, arXiv:1309.3412 [hep-th];
F.~G.~Pedro and A.~Westphal, arXiv:1309.3413 [hep-th].
\bibitem{l-22-40}
J.~A.~Adams, B.~Cresswell and R.~Easther, Phys.\ Rev.\ D\ {\bf 64}, 123514 
(2001); L.~Covi, J.~Hamann, A.~Melchiorri, A.~Slosar and I.~Sorbera, Phys.\ 
Rev.\ D\ {\bf 74}, 083509 (2006); J.~Hamann, L.~Covi, A.~Melchiorri and A.~Slosar, 
Phys.\ Rev.\ D\ {\bf 76}, 023503 (2007); M.~J.~Mortonson, C.~Dvorkin, 
H.~V.~Peiris and W.~Hu, Phys.\ Rev.\ D\ {\bf 79}, 103519 (2009); M.~Joy, 
V.~Sahni and A.~A.~Starobinsky, Phys.\ Rev.\ D\ {\bf 77}, 023514 (2008); M.~Joy, 
A.~Shafieloo, V.~Sahni and A.~A.~Starobinsky, JCAP {\bf 0906}, 028 (2009).
\bibitem{hazra-2010}
D.~K.~Hazra, M.~Aich, R.~K.~Jain, L.~Sriramkumar and T.~Souradeep, JCAP 
{\bf 1010}, 008 (2010).
\bibitem{benetti-2011}
M.~Benetti, M.~Lattanzi, E.~Calabrese and A.~Melchiorri, Phys.\ Rev.\ D\ 
{\bf 84}, 063509 (2011).
\bibitem{benetti-2013}
M.~Benetti, arXiv:1308.6406 [astro-ph.CO].
\bibitem{ci}
A.~Ashoorioon and A.~Krause, arXiv:hep-th/0607001; A.~Ashoorioon, A.~Krause 
and K.~Turzynski, JCAP {\bf 0902}, 014 (2009).
\bibitem{pso}
J.~Martin and C.~Ringeval, Phys.\ Rev.\ D\ {\bf 69}, 083515 (2004);
Phys.\ Rev.\ D\ {\bf 69}, 127303 (2004); JCAP {\bf 0501}, 007 (2005);
M.~Zarei, Phys.\ Rev.\ D\ {\bf 78}, 123502 (2008).
\bibitem{pahud-2009}
C.~Pahud, M.~Kamionkowski and A.~R.~Liddle, Phys.\ Rev.\ D\ {\bf 79}, 
083503 (2009).
\bibitem{flauger-2010}
R.~Flauger, L.~McAllister, E.~Pajer, A.~Westphal and G.~Xu, JCAP {\bf 1006}, 
009 (2010). 
\bibitem{kobayashi-2011}
T.~Kobayashi and F.~Takahashi, JCAP {\bf 1101}, 026 (2011).
\bibitem{aich-2013}
M.~Aich, D.~K.~Hazra, L.~Sriramkumar and T.~Souradeep, Phys.\ Rev. D\
{\bf 87}, 083526 (2013).
\bibitem{easther-2013}
H.~Peiris, R.~Easther and R.~Flauger, arXiv:1303.2616 [astro-ph.CO];
R.~Easther and R.~Flauger, arXiv:1308.3736 [astro-ph.CO].
\bibitem{meerburg-2013} 
P.~D.~Meerburg, D.~N.~Spergel and B.~D.~Wandelt, arXiv:1308.3704 [astro-ph.CO];
P.~D.~Meerburg and D.~N.~Spergel, arXiv:1308.3705 [astro-ph.CO].
\bibitem{ne-ps}
D.~S.~Salopek, J.~R.~Bond and J.~M.~Bardeen, Phys.\ Rev.\ D\ {\bf 40}, 1753 
(1989); C.~Ringeval, Lect.\ Notes Phys.\ {\bf 738}, 243 (2008).
\bibitem{ng-ne} 
X.~Chen, R.~Easther and E.~A.~Lim, JCAP {\bf 0706}, 023 (2007); JCAP 
{\bf 0804}, 010 (2008).
\bibitem{ng-f} 
S.~Hotchkiss and S.~Sarkar, JCAP {\bf 1005}, 024 (2010); S.~Hannestad, 
T.~Haugbolle, P.~R.~Jarnhus and M.~S.~Sloth, JCAP {\bf 1006}, 001 (2010); 
R.~Flauger and E.~Pajer, JCAP {\bf 1101}, 017 (2011); P.~Adshead, W.~Hu, 
C.~Dvorkin and H.~V.~Peiris, Phys.\ Rev.\ D\ {\bf 84}, 043519 (2011); 
X.~Chen, JCAP {\bf 1201}, 038 (2012); P.~Adshead, W.~Hu and V.~Miranda, 
Phys.\ Rev.\ D\ {\bf 88}, 023507 (2013); J.~S.~Horner and C.~R. Contaldi,
arXiv:1311.3224 [astro-ph.CO].
\end{thebibliography}
\end{document}